\documentclass[%
 reprint,
 amsmath,amssymb,
 aps,
]{revtex4-2}

\usepackage{graphicx}
\usepackage{dcolumn}	
\usepackage{hyperref}
\usepackage{bm}
\usepackage{wasysym}
\usepackage[utf8]{inputenc}
\usepackage[T1]{fontenc}
\usepackage{etoolbox}
\usepackage{titlesec}
\usepackage{subcaption}
\usepackage[dvipsnames]{xcolor}
\usepackage{amsmath}
\usepackage{dsfont}
\usepackage{indentfirst}
\usepackage{xcolor}
\usepackage{amssymb}
\usepackage{mathtools}
\usepackage{bbold}
\usepackage{accents}
\usepackage{multirow}
\usepackage{palatino} 
\usepackage{lmodern}
\usepackage{soul}
\usepackage{svg}
\usepackage{aas_macros}
\usepackage[margin=0.5in]{geometry}

\definecolor{burntsienna}{rgb}{0.91, 0.45, 0.32}

\usepackage{textcomp} 
\usepackage{amstext} 

\begin{document}

\title{Pixel domain implementation of the Minimally Informed CMB MAp foreground Cleaning (MICMAC) method}

\author{Magdy Morshed$^{1,4}$}
\email{morshed@apc.in2p3.fr}
\author{Arianna Rizzieri$^{1,4}$}
\email{rizzieri@apc.in2p3.fr}
\author{Clément Leloup$^{2,3}$}
\author{Josquin Errard$^{1}$}
\author{Radek Stompor$^{4,1}$}
\affiliation{$^{1}$ Université Paris Cité, CNRS, Astroparticule et Cosmologie, F-75013 Paris, France}
\affiliation{$^{2}$ Kavli Institute for the Physics and Mathematics of the Universe (Kavli IPMU, WPI), UTIAS, \\
The University of Tokyo, Kashiwa, Chiba 277-8583, Japan}
\affiliation{$^{3}$ Center for Data-Driven Discovery, Kavli IPMU (WPI), UTIAS, The University of Tokyo, Kashiwa, Chiba 277-8583, Japan}
\affiliation{$^{4}$ CNRS-UCB International Research Laboratory, Centre Pierre Bin\'{e}truy, IRL2007, CPB-IN2P3, Berkeley, US}
\date{\today}

\begin{abstract}
    High fidelity separation of astrophysical foreground contributions from the cosmic microwave background (CMB) signal has been recognized as one of the main challenges of modern CMB data analysis, and one which needs to be addressed in a robust way to ensure that the next generation of CMB polarization experiments lives up to its promise.
    In this work we consider the non-parametric maximum likelihood CMB cleaning approach recently proposed by some of the authors which has been shown to match the performance of standard parametric techniques for simple foreground models, while superseding it in cases where the foregrounds do not exhibit a simple frequency dependence.
    We present a new implementation of the method in pixel space, extending its functionalities to account for spatial variability of the properties of the foregrounds.
    We describe the algorithmic details of our approach and its validation against the original code as well as the parametric method for various experimental set-ups and different models of the foreground components.
    We argue that the method provides a compelling alternative to other state-of-the-art techniques.
\end{abstract}

\maketitle

\section{Introduction}
\label{intro}

The observation and characterization of the polarized cosmic microwave background (CMB) primordial anisotropies is an exciting new frontier in observational cosmology. These anisotropies on the large angular scales are a unique source of information on the fundamental mechanisms that took place in the early Universe, such as the hypothetical era of cosmic inflation. The primordial gravitational waves are expected to be generated during this process and should leave a characteristic polarization pattern on the last scattering surface in the form of $B$ modes. Their detection is one of the main scientific goals of the current and next generation CMB observatories, including BICEP~\cite{BICEP:2021xfz}, the Simons Observatory~\cite{SO_forecast}, CMB-S4~\cite{CMB-S4:2020lpa}, and LiteBIRD~\cite{LiteBIRD:2023zmo}.

Diffuse polarized astrophysical foregrounds are recognized as a major limitation in reaching the cosmological signal. Today we know that emissions due to interstellar dust and synchrotron present in our Galaxy are brighter than the primordial CMB $B$ modes in all regions of the sky across the entire range of relevant, observational frequencies. Their characterization and removal, the so-called component separation procedure, are consequently at the heart of any contemporary data analysis pipeline. There exist today plenty of component separation methods, including for instance, \texttt{FGBuster}~\cite{Stompor_2009},\cite{FGBuster}, \texttt{SMICA}~\cite{Delabrouille:2002kz, Cardoso:2008qt}, \texttt{NILC}~\cite{2009A&A...493..835D, 2012MNRAS.419.1163B, 2013MNRAS.435...18B}, \texttt{GNILC}~\cite{2011MNRAS.418..467R}, \texttt{Commander}~\cite{2004Eriksen, Eriksen:2007mx, Seljebotn_2019, Cosmoglobe}, and \texttt{B-SeCRET}~\cite{delaHoz:2020ggq}, which make different assumptions about Galactic foregrounds and/or the statistical properties of the CMB, and adopt different mathematical frameworks. The choice of the framework is of importance as different choices offer different flexibility in accommodating other effects such as the observational noise or instrumental systematics.
The amplitude of the primordial $B$-mode signal is quantified via the tensor-to-scalar ratio parameter $r$, defined as the ratio of the power spectra of the primordial tensor and scalar perturbations at wavelength $k = 0.05 \ \texttt{Mpc}^{-1}$. It turns out to be much lower than the amplitude of the already observed CMB anisotropies in intensity, and of the polarized $E$ modes. Indeed, the latest constraints suggest that the parameter $r$ satisfies $r < 0.032$ at 95\% C.L. \cite{Tristram_2022}. Competitive new measurements of $r$ need to overcome outstanding instrumental, numerical and statistical challenges which are driving the design of current observatories as well as their data analysis pipelines. 

There are generically two main types of component separation techniques, referred to as the parametric and non-parametric methods. The parametric methods rely on a physically motivated modeling of the foregrounds frequency scaling, while the non-parametric methods on complementary assumptions concerning the signals, foregrounds and/or CMB statistics. While the assumptions are key in determining the performance of the methods, they are often difficult to verify given the numerous uncertainties regarding foregrounds properties. 

\texttt{FGBuster}~\footnote{ \url{github.com/fgbuster/fgbuster}} is an implementation of the parametric approach, following the formalism proposed by~\cite{Stompor_2009} and based on the concept of a spectral likelihood. The performance of this approach has been successfully illustrated on realistic simulations over the past decade in the context of diverse experimental set-ups as the Tau Surveyor~\cite{Errard:2022fcm}, the Simons Observatory~\cite{SO_forecast,Wolz:2023lzb}, LiteBIRD~\cite{LiteBIRD:2022cntPTEP}, or with various other experimental set-ups as in~\cite{Errard:2015cxa}. 
Recently, Leloup et al. (2023) \cite{leloup2023nonparametric} generalized the spectral likelihood formalism to allow for a non-parametric modeling of the foregrounds frequency scaling. Provided additional assumptions on the CMB statistical properties, this approach has been shown to be more robust against complex foregrounds spectral energy densities (SEDs). However, the implementation of this method described in \cite{leloup2023nonparametric} could not handle the spatial variability of these SEDs which has been shown to be a factor limiting the performance of both parametric and non-parametric approaches, e.g.,~\cite{Errard:2018ctl}, and particularly important for the upcoming, high-sensitivity, large-sky fraction and/or deep, observational campaigns (LiteBIRD, CMB-S4). This paper describes a pixel domain reformulation and implementation of the Leloup et al. (2023)~\cite{leloup2023nonparametric} formalism proposing a new component separation framework potentially better adapted to dealing with realistic foreground emissions, including their frequency and spatial behaviors, in the context of upcoming CMB observations.

The paper is organized as follows. We introduce the formalism in Section~\ref{section:Formalism}. In Section~\ref{section:Gibbs sampling}, we discuss in depth all the steps of the Gibbs sampling approach used to get parameter estimates. The implementation is then described in Section~\ref{section:optimisation}, and in Section~\ref{section:Results} we present the results of the validation of the developed package, \texttt{MICMAC}, and provide a comparison with both a harmonic implementation of the non-parametric approach and \texttt{FGBuster} as representing parametric techniques.

\section{Formalism}
\label{section:Formalism}

In this section we present the overall formalism of the method, emphasizing the new and common features as compared to~\cite{leloup2023nonparametric}.

\subsection{Description of the data}

We follow the common assumption that the input data $\mathbf{d}$ can be described as the sum of a noise map $\mathbf{n}$, and a linear mixture of multiple components. Those components, denoted $\mathbf{s}$, usually contain the CMB sky signal and foregrounds components that are mixed together with coefficients defining a mixing matrix, $\mathbf{A}$, i.e.,
\begin{eqnarray}
    \mathbf{d} & = & \mathbf{A}\,\mathbf{s} + \mathbf{n}.
    \label{eq:dataModel0}
\end{eqnarray}
Here, the input data vector, $\mathbf{d}$, is a set of pixelized frequency maps concatenated together and including all considered Stokes parameters. The sky components vector, $\mathbf{s}$, includes both a CMB map $\mathbf{s_c}$ and the foreground maps, $\mathbf{s}_\mathbf{f}$. For $n$ considered foreground components, we have a total of $n+1$ components. 

In this paper, we consider only the $Q$ and $U$ Stokes parameters and consider the contributions from two polarized foreground components, dust and synchrotron. The component vector $\mathbf{s}$, can be arranged in many different ways, each of which can be useful depending on the context and/or application. 
Hereafter, we keep our considerations general introducing instead an operator $\mathbf{E}$ which extracts the CMB signal from the full component vector, i.e.
\begin{eqnarray}
\mathbf{s}_\mathbf{c} & \equiv & \mathbf{E}\,\mathbf{s}.
\label{eq:EmatDef}
\end{eqnarray}

Solving the component separation problem requires in general an estimation, implicit or explicit, of both the mixing matrix elements and the sky component maps. As discussed in~\cite{leloup2023nonparametric}, this is not possible without further assumptions needed in order to break the degeneracy between the mixing matrix elements and the component maps. In particular, they show that it is possible to remove efficiently foreground contributions from the CMB signal, producing thus foreground-cleaned CMB maps, while making only some rudimentary assumptions about the foregrounds, complemented by strong assumptions about the statistical properties of the CMB signal which are however physically justified and well tested. These assumptions ensure the existence of a unique solution to a modified component separation problem, imposing a specific structure on the mixing matrix, see Eq.~(19) of~\cite{leloup2023nonparametric}. This structure breaks degeneracies between parameters of the original mixing matrix and consequently renders the problem solvable. The foreground components recovered as a result of this procedure are then some, a priori unknown, linear mixtures of the physical foreground components, nevertheless the CMB component can be recovered in its pristine form. 
In the following, we denote this new mixing matrix as $\mathbf{B}$ and the redefined sky components as $\mathbf{s}$. The new data model of which we seek for a solution reads as
\begin{eqnarray}
    \mathbf{d} & = & \mathbf{B}\,\mathbf{s} + \mathbf{n}.
    \label{eq:dataModel}
\end{eqnarray}
The foreground assumptions required are that (1) the number of the different foreground components is known and smaller than the number of observed frequencies, that (2) each foreground component contributes to a sufficient number of observed frequency channels, and that (3) the model in Eq.~(\ref{eq:dataModel0}) is satisfied. The assumption on CMB is that it is a Gaussian sky-stationary process with some in principle unknown CMB spectra and vanishing mean. Leaving aside slight non-Gaussianity generated due to the secondary effects, predominantly gravitational lensing, these are the standard, generally accepted assumptions.
We adopt them throughout this work and leave the study of non-Gaussianities induced by lensing for future work.

As elaborated in~\cite{leloup2023nonparametric}, for each Stokes parameter only $n \,(n_f-n)$ elements of the mixing matrix $\mathbf{B}$ are a priori unknown, where $n$ is the number of foreground components ($n=2$ hereafter) and $n_f$ is the experiment-dependent number of available frequency channels.
This number exceeds the number of constraints which can be set given data for a single pixel, which is given by $n_f - n$. 
For the problem to be solvable we therefore need multiple sky pixels for which the same mixing matrix can be assumed. 
In order to ensure robustness of the method in practice, the number of constraints should be significantly larger than the number of the mixing matrix elements. We assume hereafter that this is always satisfied.

\subsection{Likelihood of the problem}\label{likelihood}

The redefined system in Eq.~(\ref{eq:dataModel}) is then solved via maximization of the data likelihood constructed assuming that the data noise is Gaussian in the pixel domain and described by a known noise covariance $\mathbf{N}$. The covariance of the CMB signal is denoted as $\mathbf{C}$. Based on those assumptions, we can describe the problem with the following log-likelihood,
\begin{align}
    \mathcal{S} \left( \mathbf{B}, \mathbf{s}, \mathbf{C} \right) & \equiv \; -2\, \mathrm{ln} \mathcal{L} \left( \mathbf{B}, \mathbf{s}, \mathbf{C} \right)  
     \label{eq:initial likelihood}
    \\
     = & \left( \mathbf{d} - \mathbf{B} \mathbf{s} \right)^\mathrm{T} \mathbf{N}^{-1} \left( \mathbf{d} - \mathbf{B} \mathbf{s} \right) \, + \, \mathbf{s}_\mathbf{c}^\mathrm{T} \mathbf{C}^{-1} \mathbf{s_{c}} + \mathrm{ln} \left| \mathbf{C} \right|.
    \nonumber
\end{align}
Following, the procedure described in~\cite{Stompor_2009}, we maximize the likelihood over foreground component amplitudes,
\begin{equation}
    \frac{\partial \mathcal{S}}{\partial \mathbf{s_{f}}} = 0 \ \Rightarrow \ \mathbf{s_{f}} \, = \, ( \mathbf{B}_\mathbf{f}^\mathrm{T}\mathbf{N}^{-1}\mathbf{B_{f}})^{-1} \mathbf{B}_\mathbf{f}^\mathrm{T}\mathbf{N}^{-1} \left( \mathbf{d} - \mathbf{B_{c}}\mathbf{s_{c}} \right),
\end{equation}
where we split the redefined mixing matrix $\mathbf{B}$ into the CMB and foreground parts,
\begin{eqnarray}
    \mathbf{B} & = & \big[\mathbf{B}_\mathbf{c}, \mathbf{B}_\mathbf{f}\big].
\end{eqnarray}
We then derive a profile likelihood of the likelihood in Eq.~(\ref{eq:initial likelihood}) maximized over the foreground amplitudes, which logarithm can be expressed using the Sherman-Morrison-Woodburry formula in the form,
\begin{eqnarray}
    \mathcal{S}_{\rm prof} & = &  \mathbf{d}^\mathrm{T} \mathbf{Pd} + \mathbf{s_c^{ML}}^\mathrm{T} \left( \mathbf{N_{c}} + \mathbf{C} \right)^{-1} \mathbf{s_c^{ML}} \nonumber \\ 
    & & + \left( \mathbf{s_{c}} - \mathbf{s_c^{WF}} \right)^\mathrm{T} \left( \mathbf{N_{c}}^{-1} + \mathbf{C}^{-1} \right) \left( \mathbf{s_{c}} - \mathbf{s_c^{WF}} \right) 
    \label{eq:First Likelihood}    
    \\
    & & + \ \mathrm{ln} \left| \mathbf{C} \right|,
\nonumber
\end{eqnarray}
with the noise covariance of the reconstructed CMB component $\mathbf{N_c}$, the Wiener filter maps $\mathbf{s_c^{WF}}$, the Maximum Likelihood solution maps $\mathbf{s_c^{ML}}$, and the projector $\mathbf{P}$, defined by,
\begin{eqnarray}
\begin{array}{l c l}\medskip
    \mathbf{N}_\mathbf{c} & \equiv & \mathbf{E}^\mathrm{T} \left( \mathbf{B}^\mathrm{T} \mathbf{N}^{-1} \mathbf{B} \right)^{-1} \mathbf{E} \\ \medskip
    \mathbf{s_c^{WF}} & \equiv & \left( \mathbf{N}_\mathbf{c}^{-1} + \mathbf{C}^{-1} \right)^{-1} \mathbf{N}_\mathbf{c}^{-1} \mathbf{s}_\mathbf{c}^\mathbf{ML}  \\ \medskip
    \mathbf{s_c^{ML}} & \equiv & \mathbf{E}^\mathrm{T} \left( \mathbf{B}^\mathrm{T} \mathbf{N}^{-1} \mathbf{B} \right)^{-1} \mathbf{B}^\mathrm{T} \mathbf{N}^{-1} \mathbf{d}  \\
    \mathbf{P}  & \equiv & \mathbf{N}^{-1} - \mathbf{N}^{-1}\mathbf{B} \left( \mathbf{B}^\mathrm{T}\mathbf{N}^{-1}\mathbf{B} \right)^{-1} \mathbf{B}^\mathrm{T}\mathbf{N}^{-1},
\end{array}
\label{eq:Definitions profile likelihood}
\end{eqnarray}
where $\mathbf{E}$ is the operator defined in Eq.~(\ref{eq:EmatDef}).
We note that $\mathbf{N}_\mathbf{c}$ expresses the data noise propagated to the CMB component, and as such it depends on the assumed mixing matrix.  

The likelihood in Eq.~(\ref{eq:First Likelihood}) was studied in~\cite{leloup2023nonparametric}. In particular, they show that after a marginalization over $\mathbf{s}_\mathbf{c}$, it reads,
\begin{eqnarray}
\begin{array} { l c l }\medskip
    \mathcal{S}_{spec} \left( \mathbf{B}, \textbf{C} \right)  & = & \mathbf{d}^\mathrm{T}\, \mathbf{P} \, \mathbf{d} + {\mathbf{s}_\mathbf{c}^\mathbf{ML}}^\mathrm{T} \left( \mathbf{N}_\mathbf{c} + \mathbf{C} \right)^{-1} \mathbf{s}_\mathbf{c}^\mathbf{ML} \\ 
    && + \; \mathrm{ln} \left| \mathbf{C} + \mathbf{N}_\mathbf{c} \right| - \mathrm{ln} \left| \mathbf{N}_\mathbf{c} \right|.
    \end{array}
    \label{eq:specLikeBiased}
\end{eqnarray}
This likelihood may be significantly biased and requires a special correction procedure. Were the CMB covariance to be known, we could proceed as in the parametric case~\cite{Stompor_2009} and simply drop the two logarithmic terms on the right hand side of Eq.~(\ref{eq:specLikeBiased}), as shown in \cite{leloup2023nonparametric}. However, this is not possible if the CMB power spectra are unknown and need to be determined at the same time as the mixing matrix elements. 
In such a case, we cannot neglect these terms as at least one of them depends on the CMB covariance. However, we can minimize their impact by adding a correction which makes them to nearly cancel each other for any choice of the mixing matrix elements $\mathbf{B}$ without changing their dependence on $\mathbf{C}$. This can be achieved by modifying the second logarithm in Eq.~(\ref{eq:specLikeBiased}) so the corrected likelihood reads,
\begin{align}
    \mathcal{S}_{spec}^{\rm corr} \left( \mathbf{B}, \textbf{C} \right)  & = \; \mathbf{d}^\mathrm{T}\, \mathbf{P} \, \mathbf{d} \, + \, \mathbf{s_c^{ML}}^\mathrm{T} \left( \mathbf{N_{c}} + \mathbf{C} \right)^{-1} \mathbf{s_c^{ML}} \nonumber \\ 
    & + \; \mathrm{ln} \left| \mathbf{C} + \mathbf{N_{c}} \right| - \mathrm{ln} \left| \mathbf{\Tilde{C}} + \mathbf{N_{c}} \right| \nonumber \\
     = & \; \mathcal{S}_{spec}\left( \mathbf{B}, \textbf{C} \right)\, + \, \mathrm{ln} \left| \mathbf{N_{c}} \right| - \mathrm{ln} \left| \mathbf{\Tilde{C}} + \mathbf{N_{c}} \right|,
    \label{eq:Corrected likelihood-Harm}
\end{align}
where $\mathbf{\Tilde{C}}$ is an approximate, fixed expectation of the CMB signal covariance. If the $B$-mode dependent part of $\mathbf{\Tilde{C}}$ is taken to be the covariance due to the lensing $B$-mode signal, this correction scheme is equivalent to first order in the signal-to-noise ratio to the one adopted in~\cite{leloup2023nonparametric}.

This likelihood provides the basis of the method presented in this work. 
The unknown parameters are here, the elements of the mixing matrix, $\mathbf{B}$, and the CMB power spectra, $\mathbf{C}_\ell$, used to define the CMB covariance, $\mathbf{C}$, as diagonal blocks; all of which are to be estimated simultaneously.

In this formalism, the foreground components are not described by any parametric frequency scaling. Instead, we use the elements of the redefined mixing matrix, which are fitted for, to remove the foreground contributions modeled as linear combinations of some implicit, internal templates. These internal templates are some, in general unknown, linear combinations of the physical components (in our case dust or synchrotron). They are defined as the non-CMB contribution to some arbitrarily selected frequency channel data which are used to define the matrix $\mathbf{B}$, as in~\cite{leloup2023nonparametric}. There are as many of these internal templates as the underlying physical components. In our case, there are therefore two internal templates and they are hereafter always chosen to correspond to the lowest and highest frequency channels of any given data set. This is beneficial from the point of view of the stability of the numerical operations involved in the computations but this choice also makes the internal templates to be dominated by the synchrotron and the dust for the lowest frequency and the highest frequency templates, respectively, making the procedure somewhat more physically intuitive. 

As no model is assumed to describe both dust and synchrotron, this approach is more robust to unknown spectral behavior of the foregrounds and should allow for a better retrieval of the cosmological signals even in the presence of the foregrounds with a complex spectral behavior. This has been indeed found in the study of~\cite{leloup2023nonparametric}.
That investigation was however conducted using an implementation of the method in the harmonic domain, capitalizing on a number of additional simplifying assumptions. Instead, in this work, we implement Eq.~(\ref{eq:Corrected likelihood-Harm}) in its full complexity directly in the pixel domain. Our goal is to maintain all the benefits of the original implementation while adding the possibility of accounting on spatial variability of the foreground properties as well as other features, such as inhomogeneous noise, of the input single frequency maps. This called for a number of advanced numerical solutions which we describe in the next Section. The resulting code is implemented in the \texttt{MICMAC}~\cite{MICMACgithub} software package, which has been made public prior to this paper.

\section{Implementation}
\label{section:Gibbs sampling}

\subsection{Algorithm}
We assume flat priors on all the parameters, but for the parameters describing the CMB covariance, be it either the CMB spectra or cosmological parameters, on which we impose the positive definiteness condition. We then use the likelihood in Eq.~(\ref{eq:Corrected likelihood-Harm}) to derive constraints on all of them.
The main complication we have to tackle is that an evaluation of the likelihood requires a computation of two determinants of large general matrices, $|\mathbf{C}+\mathbf{N}_\mathbf{c}|$ and $|\mathbf{\tilde C}+\mathbf{N}_\mathbf{c}|$. We address the first case by noting that for fixed values of the mixing matrix elements, the problem is the CMB power spectrum estimation and we can therefore capitalize on efficient solutions proposed in the context of the Bayesian CMB power spectrum estimation~\cite{Wandelt2004, 2004Eriksen, Larson:2006ds, Ducrocq_2022}. 
We consequently resort to Gibbs sampling as a technique employed to characterize the likelihood and (re-)introduce a (full sky) CMB map as one of the (latent) parameters to sample for. While this increases significantly the parameter space it avoids the need for the explicit calculation of the determinant, which instead is calculated implicitly when marginalizing over the map at the end of the procedure. The new likelihood then reads,
\begin{widetext}
\begin{align}
\begin{array}{l c l}\medskip
\mathcal{S}_{\rm prof}^{\rm corr} \left( \mathbf{s_{c}, B, C} \right) \; \equiv & \; \mathbf{d}^\mathrm{T} \, \mathbf{P} \, \mathbf{d} \, + \, {\mathbf{s}_\mathbf{c}^\mathbf{ML}}^\mathrm{T} \left( \mathbf{N}_\mathbf{c} + \mathbf{C} \right)^{-1} \mathbf{s}_\mathbf{c}^\mathbf{ML} + \left( \mathbf{s}_\mathbf{c} - \mathbf{s}_\mathbf{c}^\mathbf{WF} \right)^\mathrm{T} \left( \mathbf{N}_\mathbf{c}^{-1} + \mathbf{C}^{-1} \right) \left( \mathbf{s}_\mathbf{c} - \mathbf{s}_\mathbf{c}^\mathbf{WF} \right) \\
 & + \; \mathrm{ln} \left| \mathbf{C} \right| \, + \, \ln \left|\mathbf{N}_\mathbf{c} \right| - \mathrm{ln} \left| \mathbf{\Tilde{C}} + \mathbf{N}_\mathbf{c} \right|. 
 \end{array}
 \label{eq:Corrected likelihood-v1 Pixel-expanded}
\end{align}
\end{widetext}
To address the problem of the second determinant, we first recast the last two terms in the last expression as follows, 
\begin{align}
    \ln \left|\mathbf{N}_\mathbf{c} \right| - & \mathrm{ln} \left| \mathbf{\Tilde{C}} + \mathbf{N}_\mathbf{c} \right|  \; = \;\ln \left|\mathbf{N}_\mathbf{c} \, (\mathbf{\Tilde{C}} + \mathbf{N}_\mathbf{c})^{-1} \right|\nonumber \\
    & \; = \; 
    \ln \left|\mathbf{\tilde C}^{-1/2} \,  (\mathbf{\Tilde{C}}^{-1} + \mathbf{N}_{c}^{-1})^{-1}\,\mathbf{\tilde C}^{-1/2} \right|,
     \label{eq:secDet}
\end{align}
where $\mathbf{\tilde C}^{-1/2}$ denotes a symmetric square root of $\mathbf{\tilde C}$ which is also both symmetric and positive-definite. \\

We can then introduce a vector of latent variables $\boldsymbol{\eta}$ as a vector of Gaussian variables, see Section~\ref{section:eta term} for more details. They have, by construction, a zero mean and a covariance matrix defined as in Eq.~(\ref{eq:secDet}). These latent variables, when marginalized over, produce the required determinant without having any other impact. \\
The modified likelihood, which explicitly includes them, is then given by,
\begin{widetext}
\begin{align}
\begin{array}{l c l}\medskip
\mathcal{S}_{\rm prof}^{\rm corr} \left( \mathbf{s}_\mathbf{c}, \mathbf{B}, \mathbf{C}, \boldsymbol{\eta} \right) \; \equiv \; & \mathbf{d}^\mathrm{T} \, \mathbf{P} \, \mathbf{d} \, + \, {\mathbf{s}_\mathbf{c}^\mathbf{ML}}^\mathrm{T} \left( \mathbf{N}_\mathbf{c} + \mathbf{C} \right)^{-1} \mathbf{s}_\mathbf{c}^\mathbf{ML} + \left( \mathbf{s}_\mathbf{c} - \mathbf{s}_\mathbf{c}^\mathbf{WF} \right)^\mathrm{T} \left( \mathbf{N}_\mathbf{c}^{-1} + \mathbf{C}^{-1} \right) \left( \mathbf{s}_\mathbf{c} - \mathbf{s}_\mathbf{c}^\mathbf{WF} \right) \\
& 
+ \, \ln |\mathbf{C}| \, + \; \boldsymbol{\eta}^\mathrm{T} \left( \mathbf{\Tilde{C}}^{1/2} \, (\mathbf{\Tilde{C}}^{-1} + \mathbf{N}_\mathbf{c}^{-1}) \,\mathbf{\Tilde{C}}^{1/2} \right)^{-1} \boldsymbol{\eta}. \;
\end{array}
\label{eq:Corrected likelihood-v2 Pixel-expanded}
\end{align}

\end{widetext} 
This is the likelihood we implement in the software described here.
This expression is fully general and no assumption has been made up to this point about the structure of the noise covariance. In the \texttt{MICMAC} package, we currently assume that the noise covariance is diagonal. This is motivated by numerical considerations and is in line with the assumptions made in other approaches. The extension allowing for the noise correlations will be studied in future work.

\begin{figure}
    \centering
    \includegraphics[scale=.135]{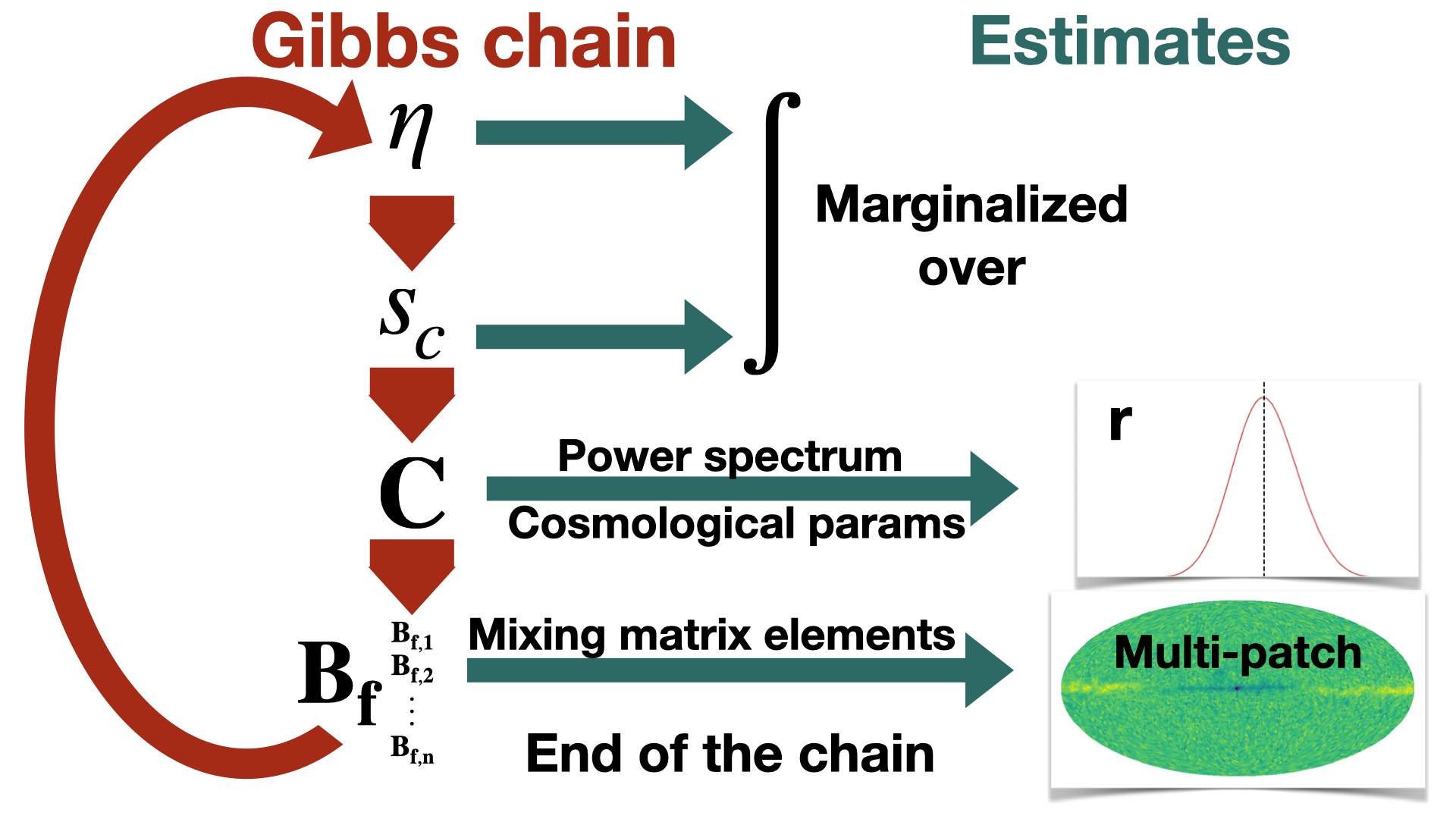}
    
    \caption{Summary of Gibbs steps and outputs of the proposed implementation: the Gaussian maps are marginalized over; the power spectrum or cosmological parameters are averaged over to obtain an estimate a posteriori; mixing matrix elements averaged over to obtain values for each patch element, see Section~\ref{section:multipatch}.}
    \label{fig:schematics_Gibbs}
\end{figure}

The Gibbs sampling scheme used for generating samples of the power spectra and the CMB map in the context of power spectrum estimation can be easily extended to add sampling of the conditional posteriors for the mixing matrix elements and for the latent variables. Our algorithm implements a Gibbs sampler for the full likelihood in Eq.~(\ref{eq:Corrected likelihood-v2 Pixel-expanded}). 
The conditional posteriors we sample from are given explicitly in Eqs.~(\ref{full_Gibbs_sampling_scheme:eta}-\ref{full_Gibbs_sampling_scheme:B_f}), and are illustrated schematically in Figure~\ref{fig:schematics_Gibbs}. 
\begin{widetext}
\begin{eqnarray} 
    \mathcal{P} (\boldsymbol{\eta}  | ....) \ \ \; & \propto & \ \boldsymbol{\eta}^\mathrm{T} \ \left( \mathbf{\Tilde{C}}^{1/2}  \ (\mathbf{\Tilde{C}}^{-1} + \mathbf{N}_\mathbf{c}^{-1}) \ \mathbf{\Tilde{C}}^{1/2} \right)^{-1} \ \boldsymbol{\eta}  
    \label{full_Gibbs_sampling_scheme:eta}\\ 
    \mathcal{P} (\mathbf{s}_\mathbf{c}  | ....) \ \  & \propto & \ (\mathbf{s}_\mathbf{c} - \mathbf{s}_\mathbf{c}^\mathbf{WF})^\mathrm{T}  \ (\mathbf{C}^{-1} + \mathbf{N}_\mathbf{c}^{-1}) \ (\mathbf{s}_\mathbf{c} - \mathbf{s}_\mathbf{c}^\mathbf{WF}) 
    \label{full_Gibbs_sampling_scheme:s_c}\\ 
    \mathcal{P} (\mathbf{C} | ....)  \ \  & \propto & \ \  \mathbf{s}_\mathbf{c}^\mathrm{T}  \mathbf{C}^{-1} \mathbf{s}_\mathbf{c} + \mathrm{ln} |\mathbf{C}|\phantom{^{X^{X^X}}}
    \label{full_Gibbs_sampling_scheme:C}\\
    \mathcal{P} (\mathbf{B}_\mathbf{f}  | ....) \, & \propto & \ -(\mathbf{d} - \mathbf{B}_\mathbf{c}\, \mathbf{s}_\mathbf{c})^\mathrm{T}  \mathbf{N}^{-1} \mathbf{B}_\mathbf{f}\, (\mathbf{B}_\mathbf{f}^\mathrm{T} \mathbf{N}^{-1} \mathbf{B}_\mathbf{f})^{-1}\, \mathbf{B}_\mathbf{f}^\mathrm{T} \mathbf{N}^{-1} (\mathbf{d} - \mathbf{B}_\mathbf{c} \,\mathbf{s}_\mathbf{c}) + \boldsymbol{\eta}^\mathrm{T} \ \mathbf{\Tilde{C}}^{-1/2}  \ (\mathbf{\Tilde{C}}^{-1} + \mathbf{N}_\mathbf{c}^{-1})^{-1} \,\mathbf{\Tilde{C}}^{-1/2} \ \boldsymbol{\eta}. 
    \label{full_Gibbs_sampling_scheme:B_f}
\end{eqnarray}
\end{widetext}

As not all the samples drawn from these conditionals can be computed directly, we need to resort to a Metropolis-Hastings sampler approach in particular for generating the samples of the mixing matrix elements using Eq.~(\ref{full_Gibbs_sampling_scheme:B_f}). This algorithm falls then into the category of the Metropolis-within-Gibbs (MwG) approaches. Such technique have been also proposed and used successfully in the context of the CMB power spectrum and parameter estimation~\cite{Jewell_2009ApJ...697..258J, 2016Racine, Ducrocq_2022}. We describe details of the implementation of this overall scheme in the following sections.

\subsection{Latent variables}
\label{section:eta term}

The latent variables introduced in Eq.~(\ref{eq:Corrected likelihood-v2 Pixel-expanded}), when marginalized over, produce the required determinant term owing to the well-known identity,
\begin{equation}
    \int e^{-\frac{1}{2} \boldsymbol{\eta}^\mathrm{T} \mathbf{A} \boldsymbol{\eta}} d\boldsymbol{\eta} = \sqrt{\frac{(2\pi)^n}{\left | \mathbf{A} \right |}}  \propto e^{-\frac{1}{2} \mathrm{ln} \left | \mathbf{A} \right |},
    \label{eq:marginalization}
\end{equation}
which holds for any symmetric definite-positive matrix $\mathbf{A}$. In our case, this matrix is given by Eq.~(\ref{eq:secDet}). 

In the proposed scheme we include these latent variables in each drawn sample and marginalize over them only at the end of the procedure. Consequently, the additional steps, which are included in the procedure are,
\begin{enumerate}
    \item For each Gibbs step, we sample the Gaussian variable $\boldsymbol{\eta}$ with the conditional probability given by Eq.~(\ref{full_Gibbs_sampling_scheme:eta});
    \item We take into account its contribution in the sample of $\mathbf{B_f}$ by adding the corresponding probability instead of the determinant term as shown in Eq.~(\ref{full_Gibbs_sampling_scheme:B_f});
    \item Once the Gibbs sampling is concluded, we marginalize over the sampled $\boldsymbol{\eta}$ variables.
\end{enumerate}

The sample of $\boldsymbol{\eta}$ on the first step above is computed as,
\begin{eqnarray} 
    \boldsymbol{\eta} & = & \mathbf{\Tilde{C}}^{1/2} \left( \mathbf{\Tilde{C}}^{-1/2} \mathbf{x} + \mathbf{N}_\mathbf{c}^{-1/2} \mathbf{y} \right),
    \label{equation_eta_prime}
\end{eqnarray}
where we draw two independent vectors, $\mathbf{x}$ and $\mathbf{y}$, composed of uncorrelated random Gaussian numbers with zero mean and unit variance for each pixel. Their respective shapes are of one component map for $\mathbf{x}$ and of all input frequency maps for $\mathbf{y}$.

The square root of the CMB noise covariance, $\mathbf{N}_\mathbf{c}^{-1/2}$, can be expressed as,
\begin{align}
\mathbf{N_c}^{-1/2} \; & =  \; \mathbf{N}_\mathbf{c}^{-1} \mathbf{N}_\mathbf{c}^{1/2} \\
 = \; & ( \mathbf{E}^\mathrm{T} (\mathbf{B}^\mathrm{T} \mathbf{N}^{-1} \mathbf{B})^{-1} \mathbf{E})^{-1} \ \mathbf{E}^\mathrm{T} (\mathbf{B}^\mathrm{T} \mathbf{N}^{-1} \mathbf{B})^{-1} \mathbf{B}^\mathrm{T} \mathbf{N}^{-1/2}, 
 \nonumber
\end{align}
so the computation of its product and the vector $\mathbf{y}$ in Eq.~(\ref{equation_eta_prime}) can be performed without the need to construct this matrix explicitly. While not a major problem in the cases studied here, this may be important in a more realistic applications whenever non-trivial pixel-pixel noise correlations are considered. We leave the exploitation of this feature for future work.

It is straightforward to verify that the computed sample has all the required properties. Indeed, the mean of the drawn samples of $\boldsymbol{\eta}$ is, 
$$ \left< \boldsymbol{\eta}  \right> =0 $$
and its covariance, 
$$ \left< \boldsymbol{\eta} \boldsymbol{\eta}^\mathrm{T} \right> = \mathbf{\Tilde{C}}^{1/2} \ (\mathbf{\Tilde{C}}^{-1} + \mathbf{N_c}^{-1}) \ \mathbf{\Tilde{C}}^{1/2}, $$
both as desired.

\subsection{Sampling of the CMB variables}

We sample the CMB variables, the sky maps $\mathbf{s}_\mathbf{c}$ and the power spectra $\mathbf{C}_\ell$ employing the techniques developed for the Bayesian CMB power spectrum estimation codes~, e.g. \cite{2004Eriksen, Wandelt2004, Jewell_2009ApJ...697..258J, 2016Racine, Ducrocq_2022}. These are performed on the second and third steps of the Gibbs sampling scheme implemented here, respectively Eq.~(\ref{full_Gibbs_sampling_scheme:s_c}) and Eq.~(\ref{full_Gibbs_sampling_scheme:C}).

The following two subsections provide more details about their implementation in the context of our approach.

\subsubsection{Sampling of $\mathbf{s}_\mathbf{c}$}

Following the conditional posterior in Eq.~(\ref{full_Gibbs_sampling_scheme:s_c}), the sample of $\mathbf{s}_\mathbf{c}$ is a Gaussian random variable with mean defined by the Wiener filter map of the CMB signal $\mathbf{s}_\mathbf{c}^\mathbf{WF}$ in Eq.~(\ref{eq:Definitions profile likelihood}) and variance given by $(\mathbf{C}^{-1}+\mathbf{N}_\mathbf{c}^{-1})^{-1}$, where both the CMB covariance and the CMB noise covariance matrix are computed for the current values of the power spectra and mixing matrix elements. To simplify the computations we assume that the CMB signal sample is always a full sky object, independently on the sky coverage of the actual data $\mathbf{s}_\mathbf{c}^ \mathbf{ML}$. The inverse noise covariance $\mathbf{N}_\mathbf{c}^{-1}$ is then extended to the full sky by padding the extra rows and columns with zeros.

We compute the mean $\mathbf{s}_\mathbf{c}^\mathbf{WF}$ by solving the second equation of Eqs.~(\ref{eq:Definitions profile likelihood}), a Wiener filter equation, using an iterative conjugate gradient (CG) solver. For this we recast the expression in Eqs.~(\ref{eq:Definitions profile likelihood}) as,
\begin{equation}
(\mathds{1} + \mathbf{C}^{1/2} \mathbf{N}_\mathbf{c}^{-1} \mathbf{C}^{1/2}) \mathbf{C}^{-1/2} \mathbf{s}_\mathbf{c}^\mathbf{WF} = \mathbf{C}^{1/2} \mathbf{N}_\mathbf{c}^{-1} \ \mathbf{s}_\mathbf{c}^\mathbf{ML}.
\label{eq:WF_CMB}
\end{equation}
To account for the missing power we add a random contribution $\boldsymbol{\zeta}$ to the mean computed by solving iteratively the following linear system of equations, 
\begin{eqnarray}
(\mathds{1} + \mathbf{C}^{1/2} \mathbf{N}_\mathbf{c}^{-1} \mathbf{C}^{1/2}) \mathbf{C}^{-1/2} \, \boldsymbol{\zeta} & = & \boldsymbol{\xi} + \mathbf{C}^{1/2} \mathbf{N}_\mathbf{c}^{-1/2} \boldsymbol{\chi}.
\label{eq:Fluct_CMB}
\end{eqnarray}
Here, $\boldsymbol{\xi}$ and $\boldsymbol{\chi}$ denote two drawn vectors of random Gaussian numbers with zero mean and unit variance, respectively of the length of one component map and of all frequency maps, and we use a CG algorithm to solve the system.

Note that the system matrix in Eq.~(\ref{eq:Fluct_CMB}) is equivalent to $(\mathbf{C}^{-1}  + \mathbf{N}_\mathbf{c}^{-1})$ in Eq.~(\ref{full_Gibbs_sampling_scheme:s_c}) as $\mathbf{C}^{-1/2} (\mathds{1} + \mathbf{C}^{1/2} \mathbf{N}_\mathbf{c}^{-1} \mathbf{C}^{1/2}) \mathbf{C}^{-1/2}=(\mathbf{C}^{-1}  + \mathbf{N}_\mathbf{c}^{-1})$. Nevertheless, the latter system is numerically more stable as noted in~\cite{2004Eriksen}.

The sample of $\mathbf{s}_\mathbf{c}$ is then given as,
\begin{eqnarray}
    \mathbf{s}_\mathbf{c} & = & \mathbf{s}_\mathbf{c}^ \mathbf{WF} \, + \, \boldsymbol{\zeta}.
\end{eqnarray}
Both the constructions of $\mathbf{s}_\mathbf{c}^\mathbf{WF}$ and $\boldsymbol{\zeta}$ ensure that $\mathbf{s}_\mathbf{c}$ is band-limited as they are defined in harmonic domain within a given range of multipoles. This allows us to control aliasing potentially occurring while performing transformations to and from the harmonic domain~\cite{libsharp}.

\subsubsection{Sampling of $\mathbf{C}_\ell$}
\label{section:C sampling}
The third step of the sampling scheme corresponds to drawing a random sample of $\mathbf{C}_\ell$ from the posterior given in Eq.~(\ref{full_Gibbs_sampling_scheme:C}), where $\mathbf{s}_\mathbf{c}$ stands for the current sample of the CMB signal. The latter is, by construction, a full sky object and we can simplify Eq.~(\ref{full_Gibbs_sampling_scheme:C}) while making the power spectra appearing explicitly,
\begin{eqnarray}
    \mathrm{P} (\mathbf{C}_\ell  | \mathbf{s}_\mathbf{c}) \propto \prod_\ell \frac{1}{\sqrt{| \mathbf{C}_\ell |^{2\ell + 1}}} \ \mathrm{exp} \left( -\frac{1}{2} \ \mathrm{tr} \ \boldsymbol{\sigma}_\ell \mathbf{C}^{-1}_\ell \right),
    \label{inverse_Wishart} 
\end{eqnarray}
where $\boldsymbol{\sigma}_\ell \equiv \sum_{m = -\ell}^\ell \mathbf{a}_{\ell m} \mathbf{a}_{\ell m}^\dag $, and $\mathbf{a}_{\ell m}$ are the spherical harmonic coefficients of ${\mathbf{s_c}}$, given as:
\begin{align}
    \mathbf{a}_{\ell m}^\mathrm{T} &= \begin{bmatrix}
           \mathbf{a}_{\mathbf{E},\ell m}^\mathrm{T}, \ \
           \mathbf{a}_{\mathbf{B},\ell m}^\mathrm{T}
         \end{bmatrix},    
    \label{eq:alm sigma_ell}
\end{align}
and $\mathbf{C}_\ell$ given by,
\begin{align}
\mathbf{C}_\ell = \begin{pmatrix}
           \mathbf{C}_\ell^\mathbf{EE} & \mathbf{C}_\ell^\mathbf{EB}\\
           \mathbf{C}_\ell^\mathbf{BE} & \mathbf{C}_\ell^\mathbf{BB}
         \end{pmatrix}.
    \label{eq:C_ell polar}
\end{align}
where $\mathbf{C}_\ell^\mathbf{EB} = \mathbf{C}_\ell^\mathbf{BE}$.\\

From this point, several approaches can be adopted. We could either sample for the power spectrum or parameterize it to sample directly for cosmological parameters. In the former case, the sample can be computed capitalizing on the fact that the posterior in Eq.~(\ref{inverse_Wishart}) can be identified as the inverse Wishart distribution for $\ell \geq 2$.

In the current implementation of $\texttt{MICMAC}$ we have chosen as the main option the parameter sampling approach and sample, as a relevant example, for the tensor-to-scalar ratio $r$ as the only unknown parameter. The sampling of the conditional posterior for $r$ is obtained from Eq.~(\ref{inverse_Wishart}) assuming that the CMB covariance $\mathbf{C}$ is given as,
\begin{eqnarray}
    \mathbf{C}(r) = \mathbf{C}^{\rm scalar} + r \ \mathbf{C}^{\rm tensor},
    \label{eq:cmbCovModel}
\end{eqnarray}
with a similar relation holding at the power spectrum level. The value of $r$ is constrained to ensure that $\mathbf{C}(r)$ is definite-positive but is otherwise arbitrary.

We assume therefore hereafter that the scalar $\mathbf{C}^{\rm scalar}$ and tensor covariance matrices, $\mathbf{C}^{\rm tensor}$, are known.
This is motivated by the fact that within the standard cosmological model, the $E$ modes power spectrum and its contamination to $B$ modes through lensing are both constrained with high precision given the current observations. For the primordial $B$-mode power spectrum, we assume that its shape is known theoretically and only its amplitude, expressed by $r$, remains to be determined. We note that we explicitly ignore here any effects which could generate non-zero $EB$ cross-correlation, such as cosmic birefringence. We restrict ourselves to $r$ as this is a case of interest to demonstrate the performance of our method and compare it with other approaches, while the code can be straightforwardly extended to add more parameters. We leave this here for future work.

The sampling of the conditional posterior for $r$ is then implemented using the MwG approach. This involves, for all other variables fixed, the following actions,
\begin{enumerate}
    \item Proposing a new sample according to a proposal distribution, here a Gaussian distribution centered on the previous sample with an arbitrarily chosen variance fixed for all steps;
    \item Computing the acceptance ratio according to the conditional probability of the random variable to sample, Eq.~(\ref{full_Gibbs_sampling_scheme:C}) assuming Eq.~(\ref{eq:cmbCovModel});
    \item Sampling a random uniform variable and determining if the proposed sample is accepted or not, otherwise setting the new sample with the value of the current one.
\end{enumerate}
This procedure can be straightforwardly generalized to sample any extended set of cosmological parameters. We note that other samplers could also be implemented in this context as, for instance, a Hamiltonian-within-Gibbs~\cite{neal2012bayesian, HMC_Taylor_Hobson} approach.

We note that sampling the power spectrum using the Inverse Wishart distribution is also proposed in the \texttt{MICMAC} package, but is not used in this work and will be discussed in future work.

\subsection{Sampling for the mixing matrix elements $\mathbf{B_f}$}
\label{section:Bf sampling}
The conditional probability of the fourth step of the Gibbs sampling scheme is given by Eq.~(\ref{full_Gibbs_sampling_scheme:B_f}).
Due to the complex form of this probability, we cannot sample directly from it and instead we use a succession of MwG approaches to sample from the conditional posteriors for each of the mixing matrix elements. This follows the same steps as described above in the context of the sampling of $r$, and the mixing matrix elements samples are drawn for each element separately in sequence. The overall scheme is represented in Fig.~\ref{fig:schematics_Gibbs}. 
Alternately we could in principle draw a sample of all the mixing matrix elements at the same time using a single MwG step. 

We expect that the performance of this latter scheme could be hampered by a substantially lower acceptance rate as optimizing such a single step approach would require some knowledge of the full covariance matrix between the $\mathbf{B}_\mathbf{f}$ elements or relying on some more advanced variants of the samplers, as e.g. on-the-fly automated step-size adjustments~\cite{Heikki_Step-size}.
Instead, in the current code, we opted for the simpler approach which has proven to be sufficient for our needs.

In the case of the so called multi-patch approach \cite{Stompor:2016hhw,Errard:2018ctl}, the implementation is different as is detailed in Section~\ref{section:multipatch}.

The first term of the conditional probability in Eq.~(\ref{full_Gibbs_sampling_scheme:B_f}) is the spectral likelihood~\cite{Stompor_2009} and is the dominant term in this conditional probability. It can be computed straightforwardly. 
The second, subdominant term involves the inversion of the following matrix,
\begin{equation}
    \mathbf{\Tilde{C}}^{1/2} (\mathbf{\Tilde{C}}^{-1} + \mathbf{N}_\mathbf{c}^{-1} ) \mathbf{\Tilde{C}}^{1/2} =  (\mathds{1} + \mathbf{\Tilde{C}}^{1/2} \mathbf{N}_\mathbf{c}^{-1} \mathbf{\Tilde{C}}^{1/2} ).
    \label{eq:c_approx to inverse}
\end{equation}
As the sample of all $\mathbf{B}_\mathbf{f}$ elements is drawn for each element separately in the proposed scheme, we need to calculate the posterior conditioned on all the parameters for each element separately. Consequently, the inverse of the matrix in Eq.~(\ref{eq:c_approx to inverse}) has to be recomputed as many times as there are unknown elements of $\mathbf{B}_\mathbf{f}$. This turns out to be very costly even if we use an iterative CG solver to avoid explicit matrix inversions and becomes quickly prohibitive in particular for experiments with many frequency channels. This inversion has to be done $n \,(n_f-n)$ times, where $n=2$ denotes the number of foreground components and $n_f$ the number of available frequency channels. This can be a large number, in particular, for a satellite mission, e.g., \textit{LiteBIRD}~\cite{LiteBIRD:2022cntPTEP} is projected to have $15$ frequency channels, what would then lead to a large, $26$ for \textit{LiteBIRD}, inversions per single Gibbs sample of the conditional posterior Eq.~(\ref{full_Gibbs_sampling_scheme:B_f}).
To bypass this difficulty we observe first that the changes of the matrix in Eq.~(\ref{eq:c_approx to inverse}) due to changes in the mixing matrix elements are typically very small. We could then resort to inverting it only once per Gibbs step and adjusting this inverse as needed using matrix perturbation theory (see section III.2.4 of~\cite{Stewart_Sun_1990}) to account for the changes in the mixing matrix elements. This would avoid any additional matrix inversion and could be therefore calculated potentially at much lower computational cost.

In our implementation, we first define $\mathbf{A} \equiv \mathds{1} + \mathbf{\Tilde{C}}^{1/2} \mathbf{N}_\mathbf{c}^{-1} \mathbf{\Tilde{C}}^{1/2}$, where the CMB noise covariance $\mathbf{N}_\mathbf{c}$ is computed assuming the current values of the mixing matrix elements $\mathbf{B}_\mathbf{f}$, i.e. as sampled in the previous step. We denote the change in the inverse of $\mathbf{N}_\mathbf{c}$ due to the change of the mixing matrix as
$\Delta \mathbf{N}^{-1}_\mathbf{c}$ and write,
\begin{align}
(\mathds{1} \, +  \, \mathbf{\Tilde{C}}^{1/2} \, & (\mathbf{N}_\mathbf{c}^{-1} \, + \, \Delta \mathbf{N}_\mathbf{c}^{-1})\,\mathbf{\Tilde{C}}^{1/2})^{-1} \; = \; \nonumber\\
    = & \; (\mathbf{A} \, + \, \mathbf{\Tilde{C}}^{1/2} \, \Delta\mathbf{N}_\mathbf{c}^{-1}  \,\mathbf{\Tilde{C}}^{1/2})^{-1} 
    \label{eq:approximation_inverse} \\
    \approx & \; \mathbf{A}^{-1} \, - \, \mathbf{A}^{-1} \, \mathbf{\Tilde{C}}^{1/2} \,\Delta\mathbf{N}_\mathbf{c}^{-1} \, \mathbf{\Tilde{C}}^{1/2} \, \mathbf{A}^{-1}, \nonumber
\end{align}
where the correction term does not require any additional matrix inversions.

This approximation remains true as long as $|| \mathbf{\Tilde{C}}^{1/2} \Delta \mathbf{N}_\mathbf{c}^{-1} \mathbf{\Tilde{C}}^{1/2} ||$ is sufficiently small. We can however control how big this change is by choosing appropriately the step-size for the elements of $\mathbf{B_f}$ within our MwG procedure.

In practice, we do not have access to $\mathbf{A}^{-1}$ and would like to avoid its computation due to computational time and memory issues. 
Similarly, we would also rather avoid computing the correction term in Eq.~(\ref{eq:approximation_inverse}) explicitly as it involves the matrix products which are also numerically very heavy. Instead, we capitalize here on the fact that in Eq.~(\ref{full_Gibbs_sampling_scheme:B_f}) we need only a product of the inverse of the matrix $\mathbf{A}$ and the current sample of $\boldsymbol{\eta}$.
We therefore resort to a CG algorithm to compute directly $\mathbf{A}^{-1} \boldsymbol{\eta}$, where $\boldsymbol{\eta}$ is the currently sampled value of this variable.
We then use Eq.~(\ref{eq:approximation_inverse}) to compute the update to the value of the product $\mathbf{A}^{-1} \boldsymbol{\eta}$ due to the change of the mixing matrix. In doing so, the correction term is computed via a sequence of matrix vector operations from right to left avoiding the need for the construction of the correction term in Eq.~(\ref{eq:approximation_inverse}). Once the update of $\mathbf{A}^{-1} \boldsymbol{\eta}$ is derived, it is then multiplied from the left by the transpose of $\boldsymbol{\eta}$ giving the second term of Eq.~({\ref{full_Gibbs_sampling_scheme:B_f}).

This approximation allows us to compute $(\mathds{1} + \mathbf{\Tilde{C}}^{1/2} \mathbf{N}_\mathbf{c}^{-1} \mathbf{\Tilde{C}}^{1/2})^{-1} \boldsymbol{\eta}$ using a CG only once per Gibbs iteration. Once this single CG is done, we can compute the samples for all other elements of $\mathbf{B_f}$ at low computational cost as described above. While we need to control the step size, the relevant distributions are typically very peaked and the step-sizes needed for their efficient exploration via the MwG step are typically small enough to fulfill comfortably all the necessary requirements.

This is the approach implemented in the \texttt{MICMAC} code~\cite{MICMACgithub}.
We note that while on each Gibbs step we sample each of the mixing matrix elements one-by-one, we do not update successively the product of $\mathbf{A}^{-1}\boldsymbol{\eta}$ reflecting the latest drawn sample of the mixing matrix elements but rather always update the inverse matrix computed via the CG and correct for the changes of all matrix elements sampled up to that point.

\subsection{Spatially variable foreground component frequency dependence}
\label{section:multipatch}

\subsubsection{Motivation}

As mentioned in section~\ref{likelihood}, the main motivation for implementing this method in pixel domain is the need of accounting on the spatial variability of the frequency dependence of the sky components.
Indeed, because of the underlying physical processes the frequency dependence of the foreground components is expected to vary with the direction of observation given the existing observational evidence \cite{krachmalnicoff2018s,akrami2020planck}, as well as theoretical models, e.g.~\cite{Hensley_Draine_2013ApJ...765..159D,Hensley_Draine_2017ApJ...836..179H,Martinez-Solaeche:2017mgz}.

Consequently, assuming a single mixing matrix $\mathbf{B}$ for the entire observed patch is likely to be a poor description of the sky components potentially leading to uncontrollable biases in the results of any component separation method.
Most straightforwardly such effects can be accounted on in pixel domain, as done in e.g.~\cite{errard2019characterizing,Puglisi:2021hqe,Carones:2022xzs}, while it can be more challenging in the harmonic domain, e.g.~\cite{Vacher:2022mvr,BICEPKeck:2022ixr,Wolz:2023lzb}.

In the component separation approaches based in pixel domain, we can readily introduce multiple mixing matrices for different parts of the observed sky or even single pixels. In this Section we describe the necessary modifications to permit such a generalization in the case of the discussed method.

We first recall that in our method we can never sample for a different mixing matrix element for each sky pixel independently~\cite{leloup2023nonparametric}. This would unavoidably result in more degrees of freedom than constraints. Such a limiting case can only be reached by the parametric approaches.
Instead, in our case we need to have the same mixing matrix for sufficiently large subsets of all sky pixels. This requires that the observed sky is partitioned into disjoint patches prior to the component separation and given at its onset. We then assign a mixing matrix element to each patch. These can differ from one patch to the other but has to be the same for all pixels of a given patch.
The patches could a priori have any shape and could be even different for different elements of $\mathbf{B}$. Indeed, in the parametric approach applications, it has been found~\cite{Errard:2018ctl} that having different patches for different scaling law parameters and different sky components allows for finding better trade-offs between the statistical errors and the biases in the results, and leads to significant improvement in the overall performance of the method. The appropriate sky partition in such cases is motivated by physical insights about the properties of the sky components. 
In contrast, in general the components in our case are not straightforwardly identifiable with the physical components. However, as mentioned earlier, it is possible to ensure that each of these components is nevertheless dominated by a single physical component. This can be typically achieved by an appropriate choice of the form of the $\mathbf{B}$ matrix, as discussed in the examples described in the following. Such choices are in fact also advisable from the perspective of the overall numerical stability of the method. In such cases, we may be able to follow on the procedures developed for the parametric approaches obtaining similar performance gains. Consequently, the implementation described hereafter has been devised to facilitate such functionality.
While conceptually this is certainly a more complex case to tackle on, on the implementation level we find that the additional complexity can be rather easily managed without significant impact on the method efficiency.

As compared to the algorithm described in the previous sections, there are two main extensions needed to allow for a patch-dependent mixing matrix. First, we need to keep track of the pixel dependence of the mixing matrix elements at all stages of the sampling, whenever applying it to frequency or component maps. Second, we also need to modify the sampling steps so that all the additional mixing matrix elements are efficiently sampled. We implemented this latter step such that different mixing matrix elements can be assigned to different set of the sky patches.

We will refer to this generalized technique as a multi-patch approach.

\subsubsection{Implementation}

The first three steps of the Gibbs sampling, Eqs.~(\ref{full_Gibbs_sampling_scheme:eta}),~(\ref{full_Gibbs_sampling_scheme:s_c}), and~(\ref{full_Gibbs_sampling_scheme:C}), depend on the CMB noise covariance matrix $\mathbf{N}_\mathbf{c}$ and not directly on the mixing matrix $\mathbf{B}_\mathbf{f}$. Therefore, in the multi-patch implementation we only need to amend the computation of $\mathbf{N}_\mathbf{c}$ to allow for the multiple mixing matrices. This is straightforward and we can employ the same sampling algorithms for these three steps as discussed above.

The sampling of $\mathbf{B_f}$ from Eq.~(\ref{full_Gibbs_sampling_scheme:B_f}) needs to be however modified. This can be done maintaining the overall scheme of the previous implementation which loops successively over all matrix elements drawing their samples one by one using a MwG step.
In the multi-patch extension, on each pass of the loop we draw samples of a given mixing matrix element for all mixing matrices assigned to different patches. Assuming no correlations between the set of patches this can be done in parallel rather than looping over the patches, leading to significant performance gains. Indeed, we first observe that the spectral likelihood term in the right hand side of Eq.~(\ref{full_Gibbs_sampling_scheme:B_f}}) can be represented as a product of spectral likelihoods computed for each of the patches and therefore can be computed independently as long as there are no noise correlations between the patches.
The second term can not be factorized and, for any patch under consideration, the correction term in Eq.~(\ref{eq:approximation_inverse}) should account for the changes in the mixing matrix elements performed for all the patches considered earlier in this step, i.e. after the matrix $\mathbf{A}$ has been computed. This is necessary to ensure that the overall scheme remains that of the Gibbs sampler and that the posteriors for each matrix elements are correctly conditioned. However, in the MwG technique used here, we only need to compute the difference of log-probabilities induced by the change of parameters and not the log-probabilities themselves. In general, we can write the contribution of the second term of the posterior in Eq.~(\ref{full_Gibbs_sampling_scheme:B_f})
to the acceptance ratio of a new value of the parameter as,
\begin{align}
\boldsymbol{\eta}^\mathrm{T}\,\mathbf{A'}\,\boldsymbol{\eta} \, - \,
\boldsymbol{\eta}^\mathrm{T}\,\mathbf{A}\,\boldsymbol{\eta} \, & = \nonumber \\
 = & \, \boldsymbol{\eta}^\mathrm{T} \mathbf{A}^{-1} \mathbf{\Tilde{C}}^{1/2} \Delta \mathbf{N}^{-1}_\mathbf{c} \ \mathbf{\Tilde{C}}^{1/2} \mathbf{A}^{-1} \boldsymbol{\eta}
\label{eq:acceptance1}
\end{align}
where $\mathbf{A}$ and $\mathbf{A'}$ correspond to the matrix $\mathbf{A}$ before and after the parameter change and we used Eq.~(\ref{eq:approximation_inverse}). The change in the inverse CMB noise covariance is due to the change in the considered parameter. As we assumed no correlations between patches, changing a parameter specific to a patch $\mathcal{P}$ can only change the diagonal block of $\mathbf{N}_\mathbf{c}$ corresponding to that patch, hence this is also the only non-zero block of $\Delta \mathbf{N}^{-1}_\mathbf{c}$. We can therefore rewrite Eq.~(\ref{eq:acceptance1}) as,
\begin{eqnarray}
\boldsymbol{\eta}^\mathrm{T}\,\mathbf{A'}\,\boldsymbol{\eta} & - &
\boldsymbol{\eta}^\mathrm{T}\,\mathbf{A}\,\boldsymbol{\eta} \, =  \\
& = & \sum_{p,p' \in \mathcal{P}}\; (\boldsymbol{\eta}^\mathrm{T} \mathbf{A}^{-1} \mathbf{\Tilde{C}}^{1/2})_p \, (\Delta \mathbf{N}^{-1}_\mathbf{c})_{p,p'}\, (\mathbf{\Tilde{C}}^{1/2} \mathbf{A}^{-1} \boldsymbol{\eta})_{p'}, \nonumber
\end{eqnarray}
which therefore depends only on the initial values of the elements of $\mathbf{B}_\mathbf{f}$, i.e. prior to this round of sampling, and the currently drawn parameters but only relevant to the current patch $\mathcal{P}$. This algorithm again requires only a single computation of $\mathbf{A}$ via a CG for each step of the Gibbs sampler. This has to be performed prior to the current round of sampling over $\mathbf{B}_\mathbf{f}$ and for their initial values as drawn on the previous step.

As we allow for different patches for different matrix elements, the computations have to be sequential with respect to the matrix elements. However, for a given matrix element the computations for different patches can be completely parallel. 

We also note that the proposed algorithm depends on the approximation in Eq.~(\ref{eq:approximation_inverse}). This again can be controlled by the means described earlier ensuring its sufficient accuracy. We note that the change $\Delta \mathbf{N}^{-1}_\mathbf{c}$ induced from the variation of a $\mathbf{B_f}$ element within a single patch can be controlled more easily than if the $\mathbf{B_f}$ is described by a single patch on the sky.

\section{Software package basics}
\label{section:optimisation}

The software \texttt{MICMAC}, which implements this algorithm, performs in the following way:
\begin{enumerate}
    \item it takes user-provided frequency maps $\mathbf{d}$, containing a mixture of dust and synchrotron foregrounds, noise, and CMB, the noise levels for each of these maps to build the noise covariance $\mathbf{N}$, the fixed expectation of the CMB signal covariance $\mathbf{\Tilde{C}}$, the first guesses for the power spectrum (or cosmological parameters) and for the mixing matrix elements, as well as the required CG convergence accuracy parameters with the CG tolerance, absolute tolerance, and associated maximum number of iterations;
    \item it executes the chain for a certain number of iterations;
    \item it outputs chains of mixing matrix elements and power spectra or cosmological parameters. The Gaussian variables $\boldsymbol{\eta}$ and $\mathbf{s_c}$ are marginalized over and not provided by default (though can be stored if required).
\end{enumerate}

The proposed algorithm is computationally heavy and requires careful optimization to ensure its practical feasibility. This has been performed on multiple levels in the developed software. 

The most involved part of the computations are spherical harmonic transforms (SHT) and conjugate-gradient solves (CG), see Section~\ref{section:Gibbs sampling}. These have to be performed repeatedly for each drawn Gibbs sample. We optimize the algorithm both in terms of minimizing the required number of these operations as well as by optimizing the run-time of each of them.
In particular, the number of CGs is kept to minimum by using the matrix inversion approximation described earlier. This has huge impact on the calculation efficiency in particular in the multi-patch version, as a ``naive'' approach would require a single CG per mixing matrix element on each Gibbs sample, making the code quickly computationally prohibitive.

Both SHTs and CGs are also fine-tuned to ensure the necessary precision while optimizing the required number of iterations. These are performed for SHTs in order to correct for the loss of orthogonality due to the adopted HEALPix grid~\cite{Healpix} and for the CGs to derive the accurate solution. We note that in the latter solution further gains are expected owing to the application of appropriate preconditioners~\cite{MAPPRAISER_2022A&C....3900576E, Seljebotn_2019}. This work is in progress and will be included at the later stage.

We also capitalize on efficient external libraries in packages which provide high performance low-level operators. In particular, after a careful evaluation of the possible numerical frameworks, we choose to perform the entire Gibbs sampling using the \texttt{JAX} package \cite{jax2018github, jax_paper}. \texttt{JAX} is a \texttt{Python} library developed towards high-performance computing applications. It features auto-differentiation, just-in-time compilation, automatic vectorization, parallel evaluation, and natural CPU and GPU support.

The current implementation of \texttt{MICMAC} involves \texttt{Healpy}~\cite{Healpy} to perform the SHT, which are thus done calling \texttt{C} routines from \texttt{Python}. 
It thus prevents the SHT to be computed using GPUs, but future work will address this problem to interface \texttt{MICMAC} with a SHT implementation based on \texttt{JAX}. 

The backbone of \texttt{MICMAC} is the Gibbs sampler which typically requires long chains of samples in order to reach the convergence. The sampling of $\mathbf{B_f}$ can further add to that as it employs the MwG sampling scheme with short step-sizes performed for each mixing matrix element sequentially. It is therefore highly desirable to minimize the burn-in time as much as possible. The \texttt{MICMAC} package proposes routines to provide input parameters sufficiently close to the actual values. These can be calculated either using the harmonic sampling approach implemented in an equivalent code to that of~\cite{leloup2023nonparametric} with a minimiser using \texttt{JAXopt} \cite{jaxopt_implicit_diff} or a Metropolis-Hastings method using \texttt{NumPyro} \cite{phan2019composable,bingham2019pyro}. Alternatively, a parametric code as implemented in the \texttt{FGBuster} package can be used to provide input parameters sufficiently close to the actual values. We have found these approaches to be particularly helpful in initiating the chains of the mixing matrix where they have been shown to dramatically reduce the burn-in time.

\section{Results}
\label{section:Results}

In this section we present results of the validation and demonstration of the developed \texttt{MICMAC} package. To span the range of possible observations of current interest, we consider two contrasting cases, ``satellite observation'' and ``ground-based observation''.
The satellite observation case is characterized by a large sky area coverage and a large number of frequency bands covering a broad range of frequencies. The ground-based setup assumes a more limited sky area with few frequency bands covering a limited frequency range. For definiteness, we model them loosely on the LiteBIRD specifications~\cite{LiteBIRD:2022cntPTEP} and the Simons Observatory Small Aperture Telescopes (SO SATs) set-up~\cite{SO_forecast}, respectively. Thus our choices reflect the ballpark properties of the current cutting-edge experiments of these types. We do not strive here for any detailed modeling of any of these as the purpose of this section is strictly to demonstrate the performance of the method and not that of the experimental set-ups. We leave more thorough investigation of specific experimental set-ups for future work.

The characteristics of the satellite mission case are extracted from Table 13 of~\cite{LiteBIRD:2022cntPTEP}, as implemented in~\texttt{cmbdb}~\cite{cmbdb}. In particular, we assume the frequency channels and spatially uniform white noise with the noise levels per frequency channels as shown in there. We adopt the sky mask of the Planck HFI (High Frequency Instrument)~\cite{Planck_Legacy_Archive} instrument with the sky coverage of $f_{\rm sky}\approx 0.6$. For the ground-based case, we take the frequency bands and noise levels from~\cite{SO_forecast} as implemented in~\cite{cmbdb}. For the mask we adopt a similar one to that planned for the SO SAT experiments with $f_{\rm sky}\approx 0.1$.

For these two set-ups, we perform a number of tests of the proposed method assuming progressively more complex foregrounds. 
We validate the results against the known inputs and compare them with other methods, which, depending on the case, is either the harmonic implementation of the same approach as described in~\cite{leloup2023nonparametric} or the parametric code~\texttt{FGBuster}~\cite{FGBuster}. 
Given our objectives here, the adopted foreground models are devised to validate and demonstrate features of the method and do not aim at reflecting all the potential complexity of the actual foregrounds. 
Consequently, we start from a simple foreground model assuming a parametric frequency scaling for all foreground components and not allowing for spatial variability of their properties, and we gradually relax both these assumptions in a controlled, albeit not always realistic, way. We then discuss the impact that the assumptions have on the performance of the method. 

The input foreground maps are generated from the \texttt{PySM} package \cite{Zonca_2021, Thorne_2017} as being called from the \texttt{FGBuster} package~\cite{FGBuster}. We include only two components, synchrotron and dust, since they are expected to be the main polarized foreground components. 
As these are expected to dominate at the very low and high end of frequencies, we define the mixing matrix $\mathbf{B}_\mathbf{f}$ so that the two modified components approximately correspond to the physical foregrounds as seen at the two extreme frequencies, see Section~\ref{section:Formalism}. 
These foreground components will be thus dominated by the synchrotron or dust respectively. However, by construction of the modified mixing matrix, each foreground component will not match the exact physical component it is dominated by and will have a subdominant contribution from the other physical component. The exact mixing will depend on the considered experimental setup. 
For each considered case the details of the foreground models are given below.

For the CMB signal, in all cases we follow the $\Lambda$CDM cosmological model with parameters taken from~\cite{PCP2018} and consider two values of the tensor-to-scalar ratio: $r=0$ or $r=0.01$. 
We use CAMB~\cite{Lewis:1999bs} to simulate CMB power spectra. 
We then use the \texttt{Healpy} \cite{Healpy} package to generate a CMB map realization from the initial CMB power spectrum.

We use HEALPix~\cite{Healpix} pixelization throughout and set the resolution parameter to $\texttt{nside}=64$ and consequently truncate all the signals at $\texttt{lmax}=128$. We do not include explicitly any beam convolution in order to facilitate a more direct comparison between different techniques, see e.g.~\cite{Rizzieri_2024, Seljebotn_2019} for a discussion of the main beam effects in the relevant context.

The instrumental noise assumed to be homogeneous and uncorrelated is simulated as an uncorrelated Gaussian field directly in the pixel domain. Noise correlations will be discussed in a future work.

For the sake of comparison and validation, we produce posterior distributions for the mixing matrix elements and the tensor-to-scalar ratio. 
We also estimate the level of foreground residuals in the recovered CMB maps as well as the total residuals (including both the foreground residuals and noise in the recovered CMB maps).
The foreground residuals $\mathbf{r}_\mathbf{f}$ are computed as
\begin{eqnarray}
    \mathbf{r}_\mathbf{f} (\mathbf{B}_\mathbf{f}) \ & \equiv & \ \mathbf{E}\,\left( \langle\mathbf{B}_\mathbf{f}\rangle^\mathrm{T}  \mathbf{N}^{-1}\, \langle\mathbf{B}_\mathbf{f}\rangle \right)^{-1} \langle\mathbf{B}_\mathbf{f}\rangle^\mathrm{T} \, \mathbf{N}^{-1} \mathbf{d}_\mathbf{f},
    \label{eqn:cmbSysRes}
\end{eqnarray}
where $\mathbf{d}_\mathbf{f}$ stands for the (noise free) foreground contributions to all frequency channels of a given experimental setup and $\langle \mathbf{B}_\mathbf{f}\rangle$ denotes the average of the mixing matrix elements sampled after the burn-in phase. The operator $\mathbf{E}$ picks the CMB component out of the multi-component vector as in Eq.~(\ref{eq:EmatDef}). 
As there is no CMB included in $\mathbf{d}_\mathbf{f}$, the CMB component of the results contains only the foregrounds leaked to the CMB as a result of the separation. We note that as we do not include any instrumental effects, there is no leakage of the CMB signal to the estimated foreground maps~\cite{Stivoli2010}.

As $\langle\mathbf{B}_\mathbf{f}\rangle$ is obtained for a single noise realization in the input maps, we expect $\mathbf{r}_{\rm \mathbf{f}}$ to contain both statistical and systematic foreground residuals and we would need to iterate over several realizations to distinguish between the two.

The systematic residuals in the foreground residuals term results from wrong assumptions that may have been done in the data model. As such, after averaging estimates from many noise realizations, any resultant bias on $r$ is completely driven by the systematic residuals. 
In the case of the parametric methods, this can be related to either the wrong frequency scaling adopted during the separation or spatial variability of the foregrounds that is not accounted for~\cite{Errard:2018ctl,Stompor:2016hhw}. 
In the case of the proposed method, only the second factor could be in principle relevant and we thus expect that the systematic residual should be negligible for our method in the absence of spatial variability. It should be noted that the correction term presented in Eq.~\ref{eq:Corrected likelihood-Harm} is not completely suppressed, so an intrinsic bias remains although it is expected to be small. The systematic residuals are usually retrieved as an average of many input realizations. Further extensions of the code are needed to shorten the execution time to permit such an analysis. This will be discussed in future work.
 
The total residuals include both the direct contribution of noise after component separation as well as the statistical error it induces on the reconstruction of foreground parameters. This is estimated as a difference of the recovered and true input CMB maps.

The estimate of both these residuals and their comparison will give information if the dominant source of residuals is due to the noise, as expected, or a wrong estimate of the mixing matrix.

All the applications were run on a single super computer node involving 40 cores. The initial parameters of the Gibbs sampling for $\mathbf{B_f}$ and $r$ are set randomly around the expected true values, and their convergence properties verified. 
The parameters related to the CG or the SHT are set using preliminary tests on the separate operations involved. 
In particular, the iterations related to the spherical harmonics transforms are selected to be $8$ to ensure a very good consistency with a typical map $\mathbf{x}$ and \texttt{alm2map}(\texttt{map2alm}($\mathbf{x}$)). The maximum multipole considered is taken to be $2\times \rm nside$. 
The CG parameters are taken to be a maximum of around $400$ iterations and a tolerance of $1\mathrm{e}-8$. 
The results displayed account for $2000$ to $5000$ iterations, although the actual sampling involves typically the double of these numbers. This is due to the removal of the burn-in phase and further tests related to change in the step-size of the involved MwG steps, typically taken between $10^{-3}$ and $10^{-5}$ depending on the parameter. The step-size choice is expected to be improved in the future versions of the package. 
We emphasize that the current implementation of the package is not final and will be subject to further optimizations. 
For the different applications, the current computation time per iteration obtained spans from $4 \rm s$ to $55 \rm s$ depending on our ability to feature a preconditioner for the CG involved and the number of frequencies involved. 
In particular, we have used a preconditioner in the full sky cases but are still investigating it in the cut sky cases. In particular, we expect the execution times to be significantly improved by the use of a preconditioner with the cut sky runs, and with a \texttt{JAX} spherical harmonic transform package.

\subsection{Cases I: Simple foregrounds}

We first consider a simple foreground case with well defined parametric scaling relations for both dust and synchrotron which are the same in all sky directions. For this purpose we adopt the \texttt{d0s0} model from the \texttt{PySM} package. We refer the reader to~\cite{PySM_software} for the details of the modeling.

\subsubsection{Full sky case}

\begin{figure*}
    \includegraphics[scale=.4]{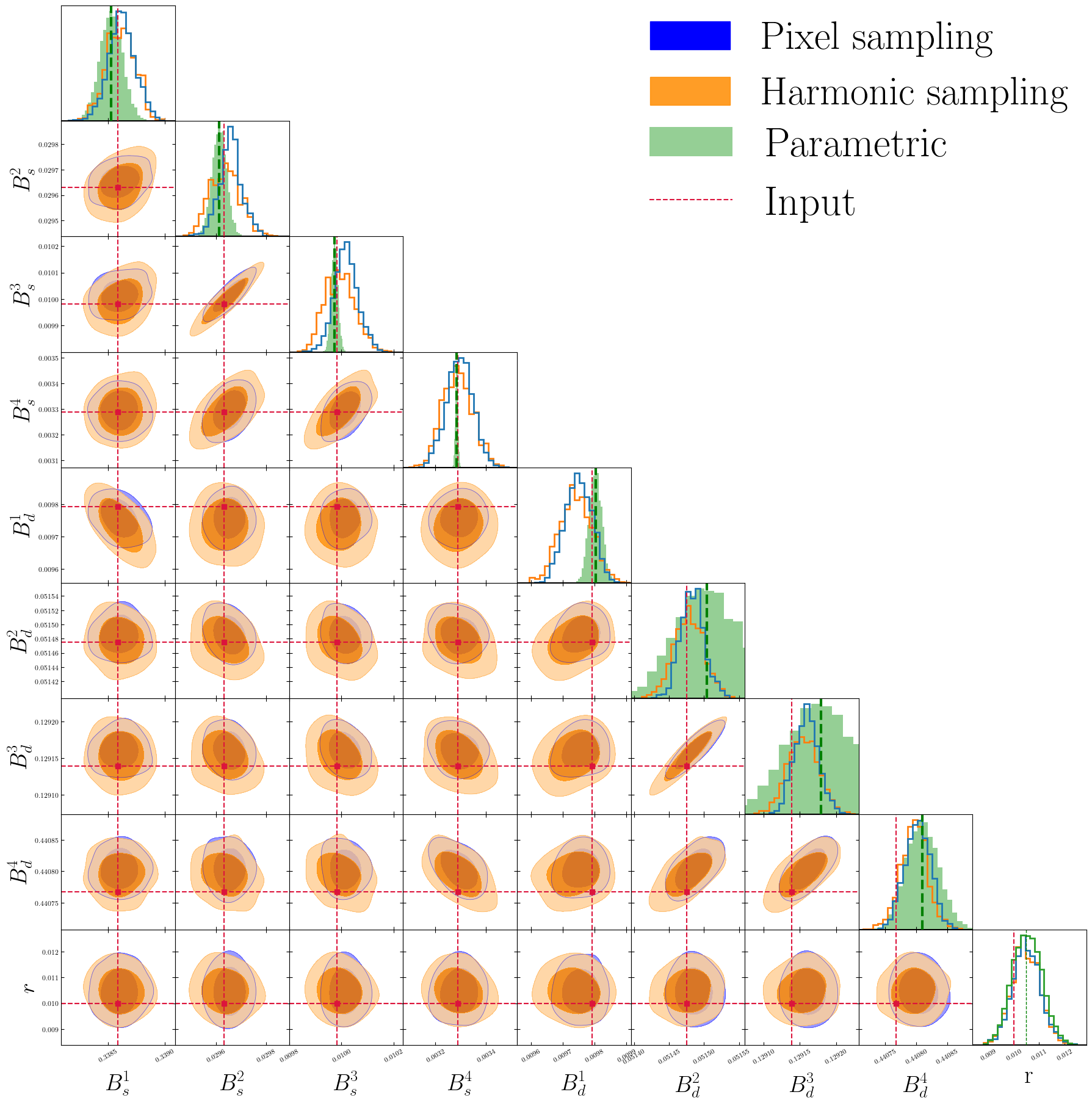}
    \caption{Corner plots showing the comparison between the pixel and harmonic implementations of the proposed method and the results from a parametric run of \texttt{FGBuster}. All the constraints are derived for the same, single realization of the input maps assuming the $\texttt{d0s0}$ model for the foregrounds, $r=0.01$, and the ground-based experiment set up with the sky coverage artificially extended to the full sky. The panels correspond to different mixing matrix elements denoted as either $B^{i}_s$ or $B^{i}_d$, corresponding to the synchrotron or dust dominated components respectively. The one-dimensional histograms obtained from the parametric approach are displayed in green. The true (noiseless) input values are plotted in red. 
    }
    \label{fig:Corner_Plot_r2_Harmonic_Pixel_SO_SAT}
\end{figure*}
\begin{figure*}
    \includegraphics[scale=.4]{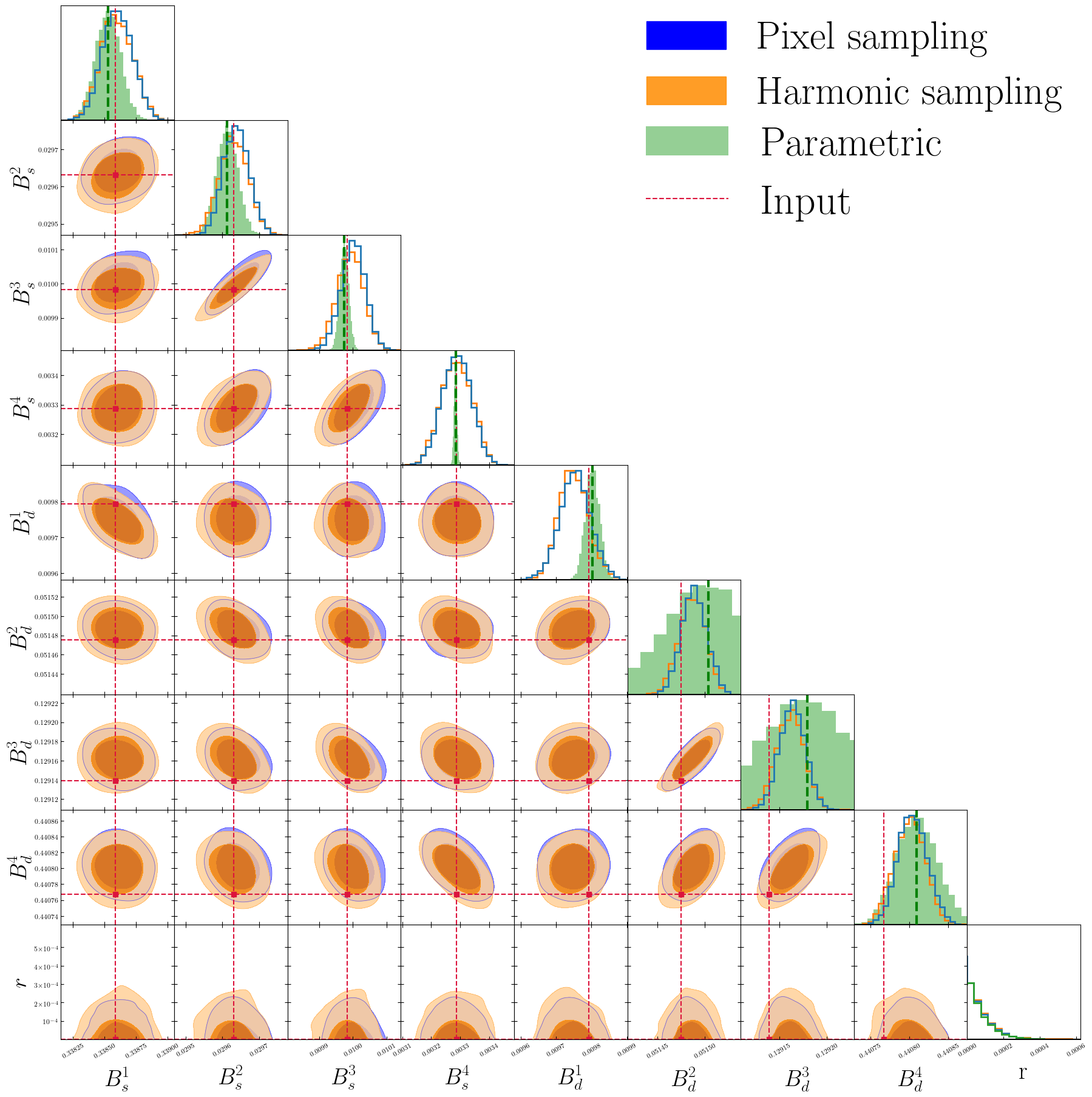}
    \caption{As Fig.~\ref{fig:Corner_Plot_r2_Harmonic_Pixel_SO_SAT} but for $r=0$. The contours on $r$ in the bottom panels appear squashed due to the condition imposed on $r$ on the MwG step, which ensure positive definiteness of $\mathbf{C}(r)$ as defined in Eq.~\ref{eq:cmbCovModel}. This choice has been assumed for concreteness here but alternative choices can be straightforwardly included.}\label{fig:Corner_Plot_r0_Harmonic_Pixel_SO_SAT}
\end{figure*}
First, we validate the new pixel domain $\texttt{MICMAC}$ code by comparing it with the harmonic implementation of the method from Eq.~(\ref{eq:Corrected likelihood-Harm}), equivalent to that of~\cite{leloup2023nonparametric}. This is done using full sky data and shown in the posteriors displayed in Figs.~\ref{fig:Corner_Plot_r2_Harmonic_Pixel_SO_SAT} and~\ref{fig:Corner_Plot_r0_Harmonic_Pixel_SO_SAT} with the frequency coverage and noise levels of the ground-based experiment. Assuming full sky, while artificial, allows us to remove any potential source of essential differences between the two codes and we expect that they should lead to consistent results. This is indeed seen in the figures which show very good agreement between the two sets of results as well as a very good agreement with the input data. 

The results in this case are derived using three different component separation approaches as mentioned earlier. The parametric approach assumes parametric scaling laws as used in the simulations and fit for three unknown parameters: spectral dust index $\beta_\mathrm{d}$, dust temperature $T_\mathrm{d}$ and synchrotron power law index $\beta_\text{s}$ assumed to be the same for all the sky pixels. For the non-parametric approaches we assume a single set of unknown mixing matrix elements, again the same for all considered sky pixels.

The results of the parametric approach come from a single run of \texttt{FGBuster} from which we derive a mixing matrix following the procedure of \cite{leloup2023nonparametric}. The $\mathbf{B_f}$ parameters are all consistent with the estimates of the harmonic and pixel sampling. They are compatible as well with the recovered values from the parametric approach. We note the parametric approach does not estimate each $\mathbf{B_f}$ parameter independently, as is done with the pixel and harmonic samplers. Instead, the estimates of the $\mathbf{B_f}$ values are computed from estimated $\{ \beta_\mathrm{d}, T_\mathrm{d}, \beta_\text{s} \}$ parameters, and as such are not independent. 
The parametric posteriors for $r$ are obtained from the CMB map extracted using the \texttt{FGBuster} mixing matrix elements, through a Metropolis-Hastings approach with the cosmological likelihood parametrized by $r$ as in Eq.~\ref{eq:cmbCovModel}. 
The true (noiseless) values of $\mathbf{B_f}$ are indicated for comparison purposes with dashed red lines. They are consistent with the recovered posteriors within the estimated scatter as are the posteriors for $r$. As the recovered posteriors are derived from a single realization of the noise we do not expect them to scatter around the true values of these parameters, as it is indeed the case. 

\begin{figure}
    \includegraphics[scale=.6]{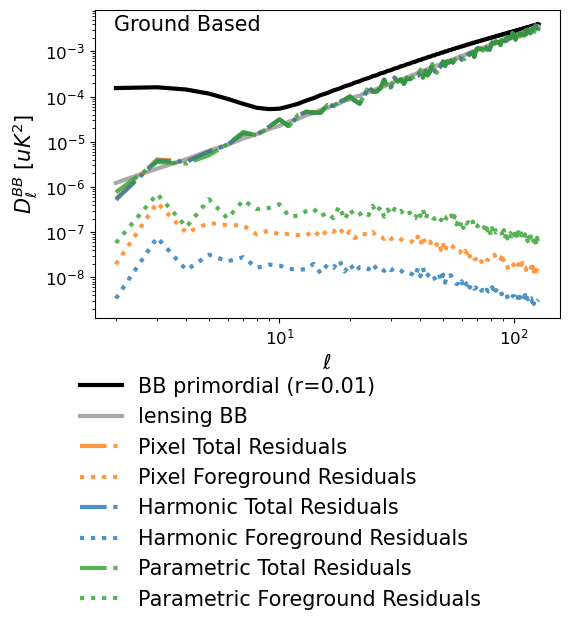}
    \caption{
    Total and foreground residuals for all methods considered here in the case of the ground experiment ex\-ten\-ded to the full sky, the \texttt{d0s0} foreground model, and $r=0.01$. 
    The foreground residuals are essentially negligible in all three cases and their amplitudes are very sensitive to the implementation and run details. 
    The noise propagated to the CMB component is roughly at the same level as the lensing BB curve, around $4.0 \rm \mu K$-$\rm arcmin$.
    }
    \label{fig:Residals_Harmonic_Pixel_SO_r2}
\end{figure}

\begin{figure}
    \includegraphics[scale=.6]{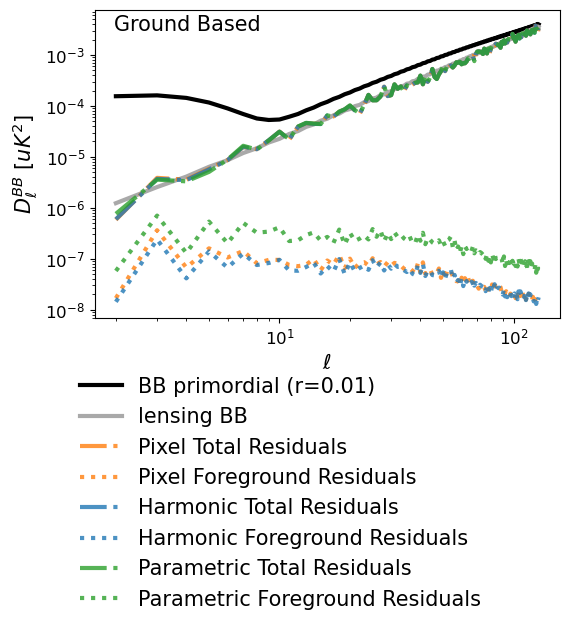}
    \caption{As Fig.~\ref{fig:Residals_Harmonic_Pixel_SO_r2} but for $r=0$. 
    }
    \label{fig:Residals_Harmonic_Pixel_SO_r0}
\end{figure}

As the mixing matrix elements are recovered in these cases with high precision the corresponding total residuals are noise dominated while the (noiseless) foreground residuals are negligible. This is illustrated in Figs.~\ref{fig:Residals_Harmonic_Pixel_SO_r2} and ~\ref{fig:Residals_Harmonic_Pixel_SO_r0}. The difference between the foreground residuals in the harmonic and pixel codes, seen in Fig.~\ref{fig:Residals_Harmonic_Pixel_SO_r2}, albeit noticeable is completely irrelevant for the final conclusions and the limits on $r$. We attribute it to the details of the implementation choices made in both codes. For instance, the harmonic and pixel codes use different sampling methods, with the harmonic sampling implementing a Metropolis-Hastings sampling and the pixel one a Gibbs sampling. 
In the harmonic approach, the $\mathbf{B_f}$ element are sampled together with the Metropolis-Hastings sampling using a covariance matrix obtained through a Fisher analysis, while in the pixel approach, the $\mathbf{B_f}$ elements are sampled one after the other without this prior knowledge with a Gibbs sampling scheme. We thus attribute the small difference between posteriors to this difference and to the fact that the results are taken with a limited number of samples. 

\begin{figure}[b]
    \includegraphics[scale=.55]{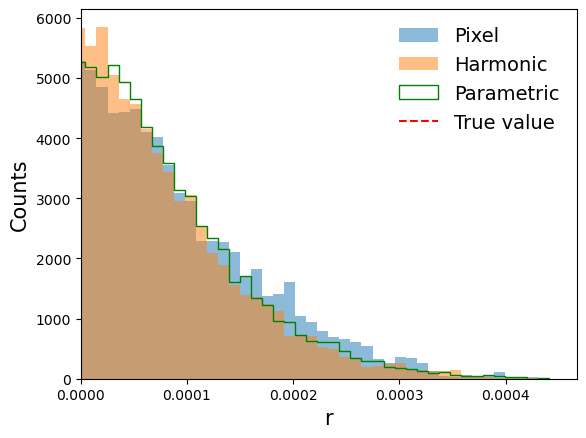}
    \caption{A comparison of the histograms of the samples of $r$ drawn in the pixel, harmonic and parametric implementations for the same single input maps and the $\texttt{d0s0}$ foreground model with $r=0$, for the satellite setup extended to the full sky.}
    \label{fig:Histogram_r0_Harmonic_Pixel_LB}
\end{figure}

\begin{figure}
    \includegraphics[width=.45\textwidth]{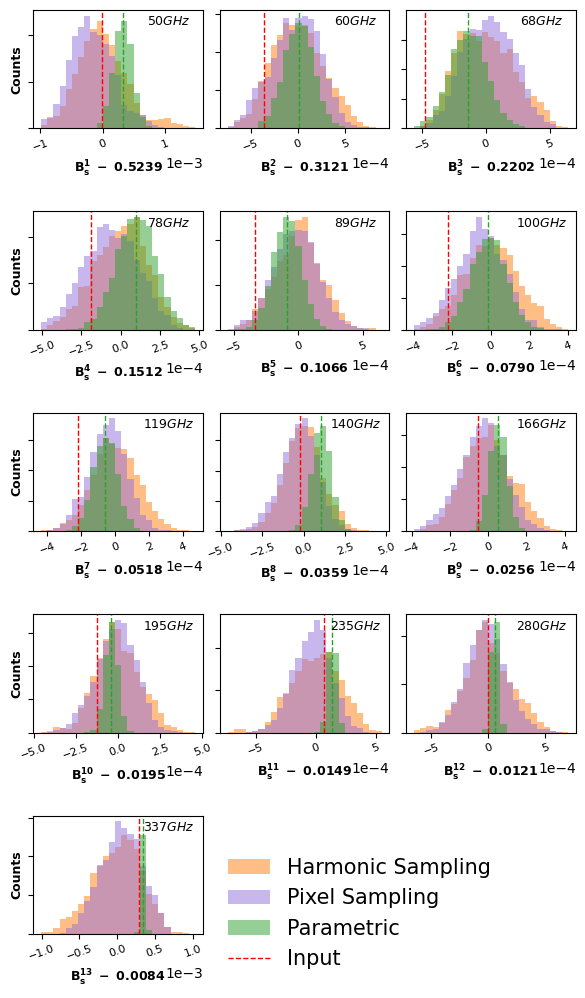}
    \caption{Comparison of the pixel and harmonic implementation of the non-parametric method for the satellite observation extended to the full sky coverage. Results of the parametric run of \texttt{FGBuster} are also plotted in green. The histograms show the 1-dimensional posteriors for the mixing matrix elements corresponding to synchrotron dominated component. They are derived assuming the $\texttt{d0s0}$ foregrounds and $r=0$. The true input values are plotted in red. 
    }
    \label{fig:Bf1_r0_Harmonic_Pixel_LB}
\end{figure}

\begin{figure}
    \includegraphics[width=.45\textwidth]{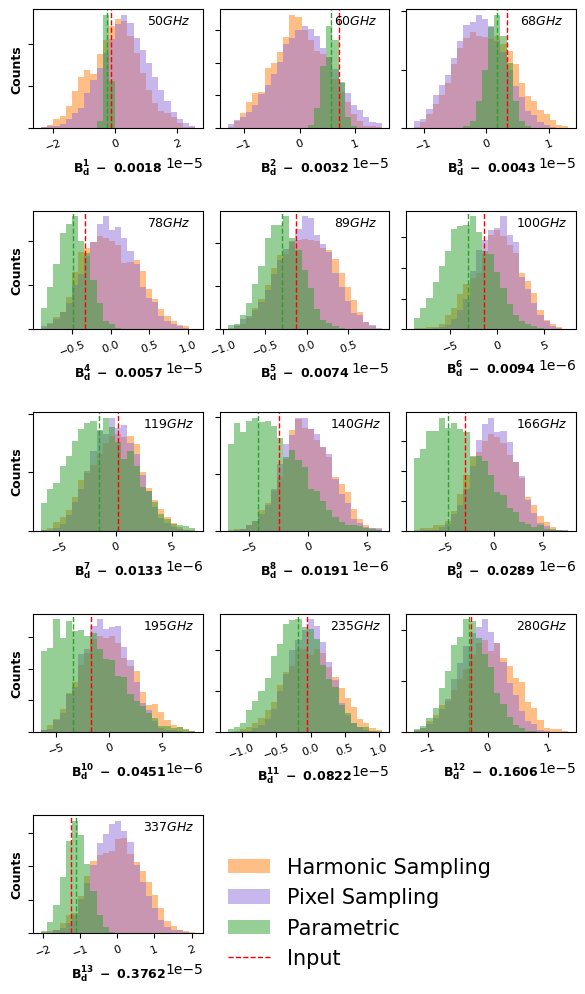}
    \caption{As Fig.~\ref{fig:Bf1_r0_Harmonic_Pixel_LB} but for the dust dominated mixing matrix coefficients.}
    \label{fig:Bf2_r0_Harmonic_Pixel_LB}
\end{figure}

The results for the satellite observation choice of frequencies and noises are similarly consistent between the harmonic and pixel sampling runs. The corresponding posteriors are displayed in Figs.~\ref{fig:Histogram_r0_Harmonic_Pixel_LB},~\ref{fig:Bf1_r0_Harmonic_Pixel_LB} and ~\ref{fig:Bf2_r0_Harmonic_Pixel_LB}. 
The histograms from the parametric run are computed using the scaling laws assumed for each of the components and the samples of the spectral parameters, $\{ \beta_\mathrm{d}, T_\mathrm{d}, \beta_\text{s} \}$. 
The parametric posterior for $r$ is again computed from a Metropolis-Hastings approach using a cosmological likelihood as in Eq.~\ref{eq:cmbCovModel} and applied to the CMB separated map from the parametric estimates of the mixing matrix elements. 
As in the previous figure, the input parameters are shown with red dashed lines. We can see that for the non-parametric results the agreement is very good throughout, while there is potential tension for some of the mixing matrix parameters in the parametric cases. 
We emphasize again the input values for the mixing matrix elements are not the ground truth which must be retrieved by the component separation methods employed, as the results are shown only for one noise realization. 
We note that while the spectral parameter estimates themselves agree rather well with their true values, the best fit value of the synchrotron index is only marginally consistent with the true value within $95$\% C.L. This in turn leads to the tensions seen in the histograms in Fig.~\ref{fig:Bf1_r0_Harmonic_Pixel_LB}. As all tensions seen in the figure are traceable to the same parameter, they are strongly correlated, and their statistical significance is lower than we would naively think just looking at the histograms. As we do not see such an effect for any other noise realization we investigated, we interpreted this tension as being merely a statistical occurrence. We point out that the spectral parameter fits in the parametric case are determined globally using simultaneously all frequency channels. This is unlike the non-parametric cases, where the fits are done for each channel separately, and it is therefore not surprising that no tensions is seen in the non-parametric results for any of the mixing matrix elements. 

\begin{figure}
    \includegraphics[scale=.6]{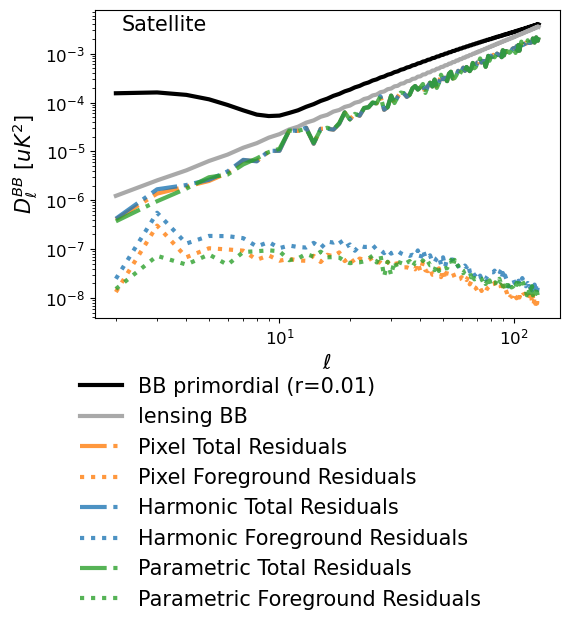}
    \caption{Residuals for the pixel and harmonic implementation of the non-parametric method for the satellite observation extended to the full sky coverage. They are derived assuming $\texttt{d0s0}$ foregrounds and $r=0$. The primordial $BB$ spectrum for $r=0.01$ is only plotted for comparison purposes.
    The noise propagated to the CMB component is expected to be below the lensing BB curve, the former being at a level of $3.0 \rm \mu K$-$\rm arcmin$ and the latter at $4.0 \rm \mu K$-$\rm arcmin$.}
    \label{fig:Residals_Harmonic_Pixel_LB}
\end{figure}

This interpretation is further supported by the fact that the residuals, both the noise-dominated total residuals and the noiseless foreground residuals, as shown in Fig.~\ref{fig:Residals_Harmonic_Pixel_LB} and obtained with the different methods are indeed fully consistent. 
Consequently, all the methods also lead to consistent results on the CMB signal, validating our approach.



\subsubsection{Cut-sky case}

\begin{figure}[!t]
    \includegraphics[scale=.55]{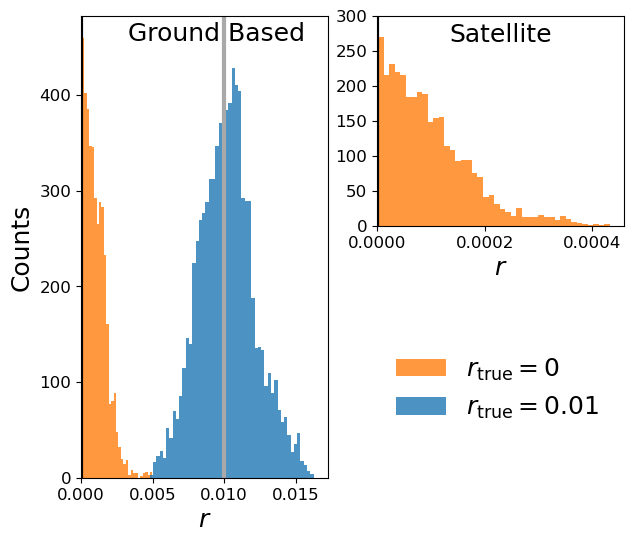}
    \caption{Histogram of drawn samples of $r$ for the ground-based (left) and satellite (right) experiments for $r=0$ (orange) and $r=0.01$ (blue). These have been obtained using the \texttt{MICMAC} pixel domain code for the \texttt{d0s0} foreground model and assuming the same mixing matrix for all sky pixels.
    }\label{fig:Histo_cutsky_r}
\end{figure}

We subsequently apply the \texttt{MICMAC} pixel domain code to the satellite and ground-based observations, as defined earlier, thus including their respective, and partial, sky coverage. The results are qualitatively consistent with those discussed earlier so we do not show them here and we focus directly on the constraints on $r$, see Fig.~\ref{fig:Histo_cutsky_r}. For the ground set-up and the case with $r= 0.01$, the estimated best-fit value 
is within $\sim 5\%$ of the true value corresponding to $\sim 10$\% of the estimated uncertainty within $95$\% C.L.. The latter is found here to be roughly $\sim 3\times10^{-3}$.

In the case with no primordial gravitational waves, $r=0$, our results are fully consistent with no detection of $r$ with the upper limits at $\sim 2\times 10^{-3}$ and $\sim 2\times10^{-4}$ ($95$\% C.L.), for the ground-based and space experiments, respectively. 

\begin{figure*}
    \includegraphics[scale=.5]{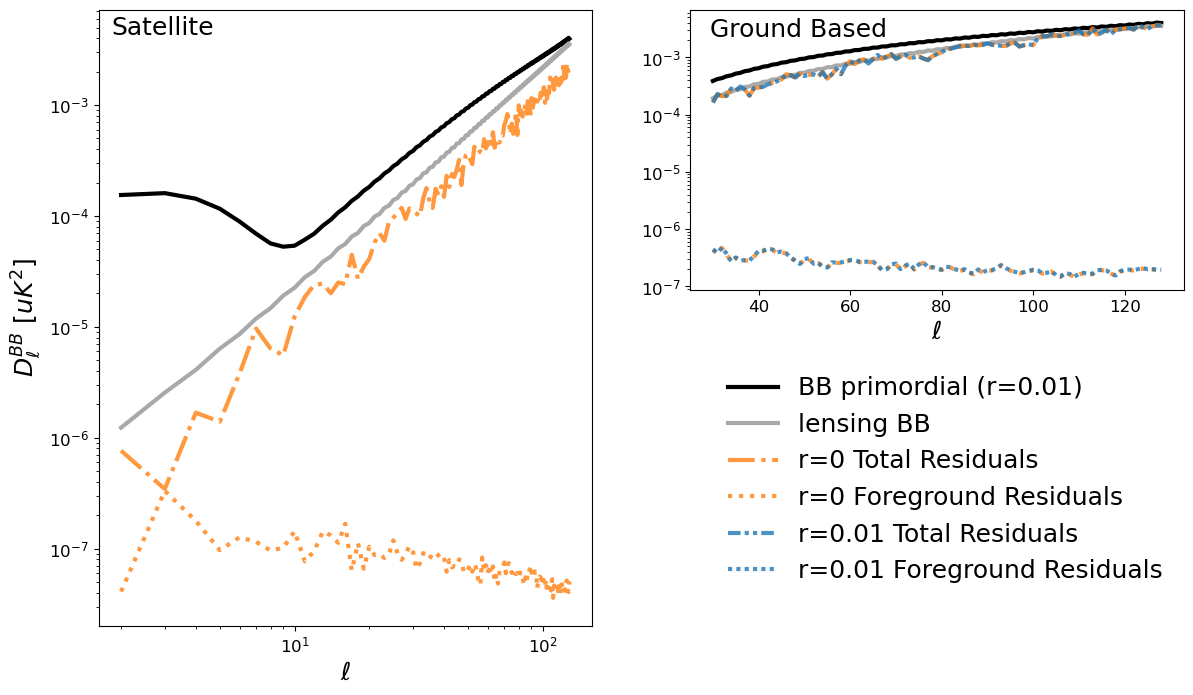}
    \caption{Residuals for the ground-based and satellite observations considered here assuming the $\texttt{d0s0}$ foregrounds, for a case with $r=0$ for both observations and and an additional case $r=0.01$ for the ground-based observation displayed with blue lines. The $\ell$ range for the ground-based experiment is limited to $\ell_{\rm min}= 30$ due to the observed sky area. These results have been derived using the \texttt{MICMAC} code assuming a single mixing matrix for all the pixels of the corresponding observed sky area. 
    }
    \label{fig:Residuals_cutsky}
\end{figure*}

We also show the corresponding spectra of the total residuals and the foreground residuals in Fig.~\ref{fig:Residuals_cutsky}. As expected, the former is completely dominated by the noise and the latter completely subdominant. This is fully consistent with the absence of any bias on the recovered value of $r$ as discussed earlier.

\subsection{Cases II: Foregrounds without simple parametrization of the scaling laws}\label{section:fgs_intermediate_complexity1}

The second main validation of the \texttt{MICMAC} code is performed on foreground models without spatial variability but which have SEDs which are not straightforwardly parametrizable. 

The model chosen is a customized version of the \texttt{d7} and \texttt{s1} models of \texttt{PySM}, see~\cite{PySM_software}. The \texttt{d7} dust model is more complex as it not only does not have a parametric SED but also both \texttt{d7} and \texttt{s1} feature spatial variability. 
To remove their spatial variability, we average the frequency scalings over pixels and use them to compute the foreground contributions to all required frequency channels. 

This procedure ensures that the frequency scaling are not readily parametric and that they are not pixel-dependent. 
As the starting model is that of \texttt{d7s1}, we refer to these simulations as \texttt{\textit{customized d7s1 without spatial variability}}. 
In this configuration and all the latter, the foregrounds considered are non-Gaussian. This also allows us to emphasize that the proposed method does not assume any foreground Gaussianity. 

\begin{figure}
    \includegraphics[scale=.55]{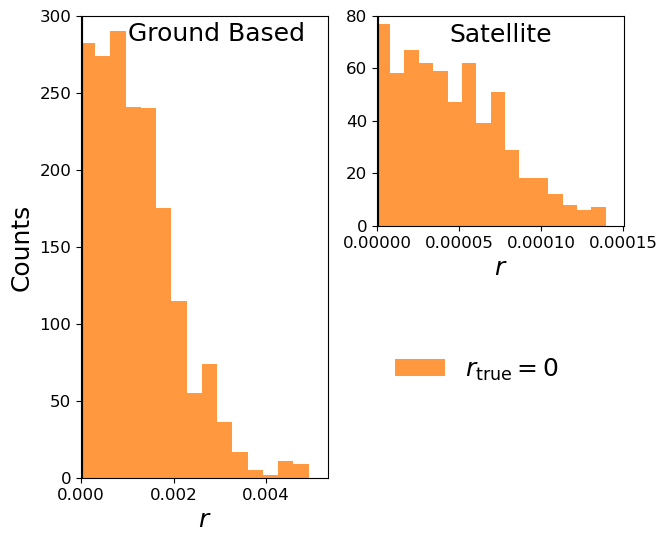}
    \caption{Histogram of samples of $r$ for the ground-based and satellite observations considered with $r=0$ assuming the \texttt{\textit{customized d7s1 without spatial variability}} model. 
    These results have been derived using the \texttt{MICMAC} code assuming a single mixing matrix for all the pixels of each observed sky area.
    }
    \label{fig:No spv - Customized d7s1 -- Histo r}
\end{figure}
\begin{figure*}
    \includegraphics[scale=.55]{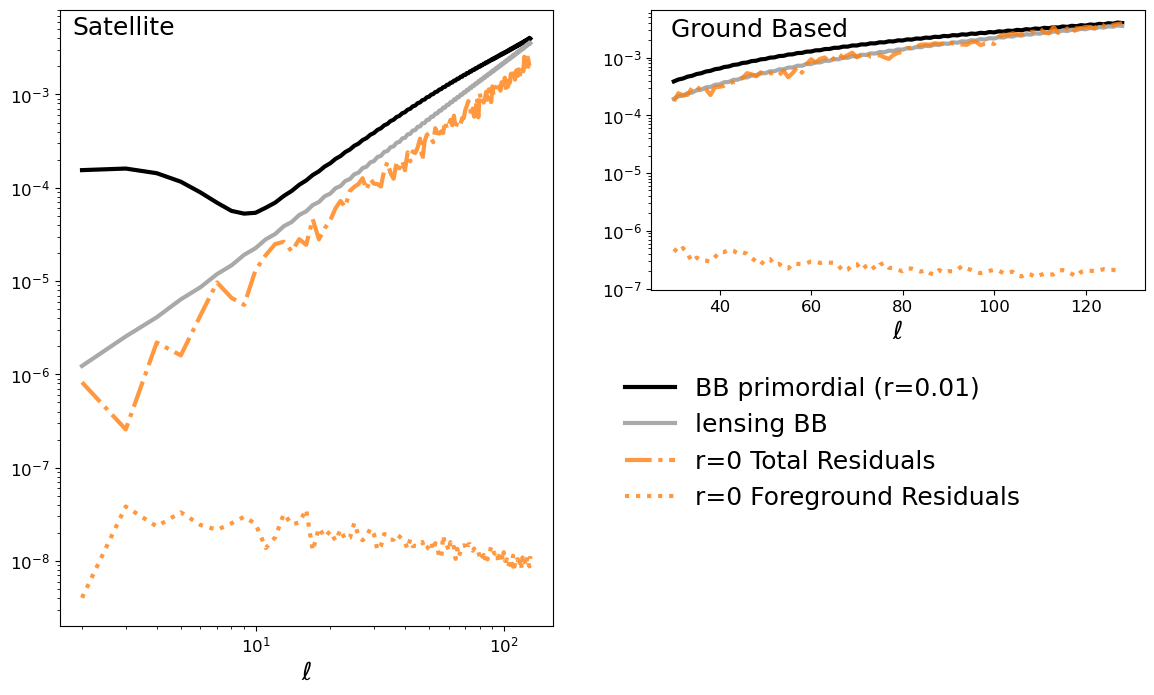}
    \caption{As Figure~\ref{fig:Residuals_cutsky} but with the \texttt{\textit{customized d7s1 without spatial variability}} model  and considering $r=0$ only.
    }
    \label{fig:No spv - Customized d7s1 -- Residuals}
\end{figure*}
We validate the pixel sampling using these foreground maps with both ground-based and satellite experiment cases their corresponding, and partial sky, coverage and $r=0$. The results are shown in Figs.~\ref{fig:No spv - Customized d7s1 -- Histo r} and~\ref{fig:No spv - Customized d7s1 -- Residuals}. The histograms of $r$ are in a very good agreement with the input values and we obtain upper limits at $\sim 2\times 10^{-3}$ on the ($95$\% C.L.) for the ground-based experiment, and $\sim 2\times10^{-4}$ ($95$\% C.L.) for the case of the satellite. The foreground residuals are completely subdominant compared to the total residuals as it was also the case for \texttt{d0s0}. 

The next section addresses the ability of the \texttt{MICMAC} package to handle spatial variability of the component scaling laws.

\subsection{Cases III: Foregrounds with spatially variable frequency scaling}

\subsubsection{Case of parametric SEDs}
\label{section:multipatch_results}

In this section we consider foregrounds with parametric but spatially varying frequency scalings.
As explained in Section~\ref{section:multipatch}, the main interest of having the pixel-based implementation of the method as compared to the harmonic implementation is in its potential ability to account for spatial dependence of foreground frequency scaling.

To demonstrate the actual performance of our multi-patch implementation as described in Section~\ref{section:multipatch} we create foreground simulations based on a customized version of the \texttt{PySM} models \texttt{s1} (for synchrotron emission) and \texttt{d1} (for thermal dust emission).
These models feature spatially varying spectral indices for both spectral indices, the synchrotron, $\mathbf{\beta}_{\mathrm{synch}}$, and the dust, $\mathbf{\beta}_{\mathrm{dust}}$, as well as the dust temperature, $\mathrm{T}_{\mathrm{dust}}$.
From those we create a new model we call \texttt{\textit{customized d1s1}}, where the spatial variability of all the spectral parameters is restricted to the angular scales larger than those given by the HEALPix pixel sizes with \texttt{nside}=1. We do that by simple averaging values of all the scaling parameters within each \texttt{nside}=1 pixel. We then use them to produce single frequency maps for the considered experimental setups. Hereafter we focus our discussion on the satellite cut-sky observation as, given its greater sky coverage, we expect the SEDs spatial variability to have a bigger role. We run \texttt{MICMAC} in two configurations: 
first, we ignore the presence of the variability and assume only a single $\mathbf{B}_\mathbf{f}$ for the entire observed sky, and then we allow for a different mixing matrix in each \texttt{nside}=1 pixel.
We thus expect to have a non negligible bias on the final recovered $r$ value in the former case, which should then be fully resolved in the latter, where the assumed model is sufficient to account for all the features of the simulations. 

\begin{figure}
    \includegraphics[scale=.65]{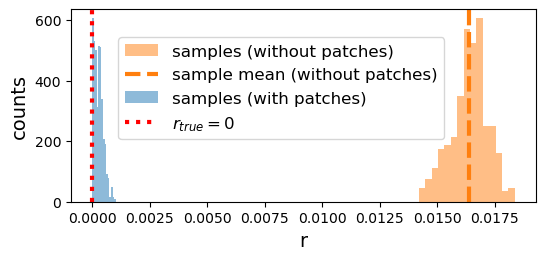}
    \caption{Histogram of the sampled values of $r$ for the \texttt{\textit{customized d1s1}} model with spatial dependence of the frequency scalings and the satellite experiment. The true input value of $r=0$ is represented by the red vertical dotted line. The samples from the \texttt{MICMAC} run with different mixing matrices assigned to different patches (blue histogram) are in good agreement with it, while the samples from the \texttt{MICMAC} run with a single mixing matrix (in orange) display a large bias.
    The two histograms have been normalized to the same value at the peak.
    }
    \label{fig:multipatch_parametric_r}
\end{figure}

\begin{figure}
    \includegraphics[scale=0.65]{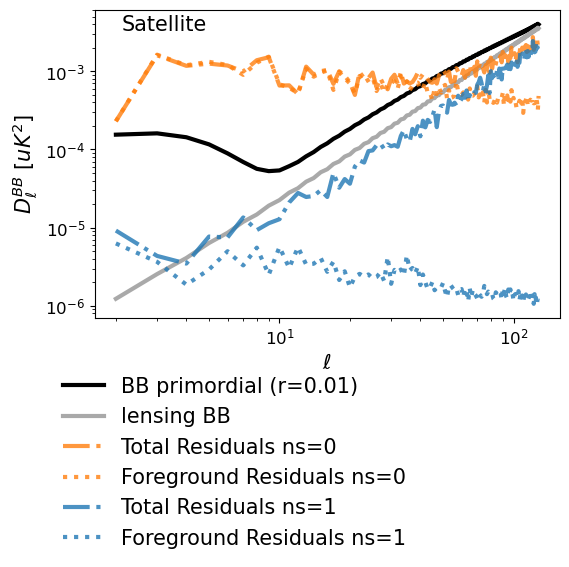}
    \caption{Residuals after component separation for the \texttt{\textit{customized d1s1}} model with spatial variability of the parametric frequency scalings. Both the foreground residuals and the total residuals are shown.
    }
    \label{fig:multipatch_parametric_Cls}
\end{figure}

We present the histograms of the recovered values of $r$ in Fig.~\ref{fig:multipatch_parametric_r}. 
The corresponding mean and standard deviation on $r$ for the run without patches in the component separation are given by $r = (16.4 \pm 1.6) \times 10^{-3}$ ($95\%$ C.L.), while for the run with patches in the component separation we obtain $r < 0.8 \times 10^{-3}$ ($95\%$ C.L.). 
In Fig.~\ref{fig:multipatch_parametric_Cls} we show the power spectra of the total and foreground residuals. The foreground residuals term is shown to be much lower in the run with multiple patches than in the run with a single one, denoted in the Figure as without patches.
All these results are fully consistent with our expectations.


\subsubsection{Case of spatially-dependent, non-parametric SEDs}

\begin{figure}[!ht]
    \includegraphics[scale=.65]{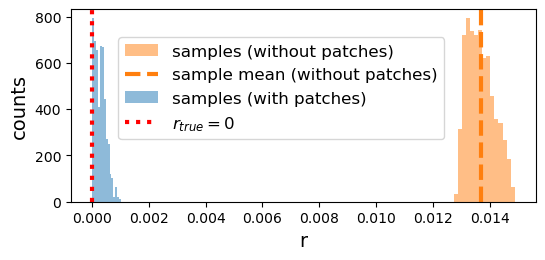}
    \caption{As Figure~\ref{fig:multipatch_parametric_r} for the \texttt{\textit{customized d7s1 with spatial variability}} model of the non-parametric frequency scaling laws. 
    }
    \label{fig:multipatch_non_parametric_r}
\end{figure}

\begin{figure}
    \includegraphics[scale=.65]{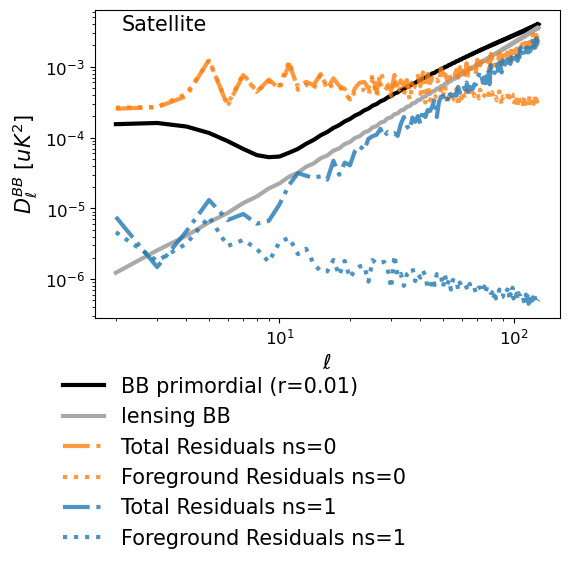}
    \caption{As Figure~\ref{fig:multipatch_non_parametric_r} for the \texttt{\textit{customized d7s1 with spatial variability}} model  of the non-parametric frequency scaling laws. 
    }
    \label{fig:multipatch_non_parametric_Cls}
\end{figure}
We now consider a more involved foreground model, which allows for foregrounds with both spatially variable and non parametric frequency scalings.
In the same spirit as what we described in Sections~\ref{section:fgs_intermediate_complexity1} and~\ref{section:multipatch_results}, we build our customized version of the \texttt{d7} and \texttt{s1} foreground models. 
The procedure we follow is similar to the one described in Section~\ref{section:fgs_intermediate_complexity1}, but we now allow for some spatial variability. Specifically, we use frequency maps produced with \texttt{PySM} \texttt{d7s1} model to produce frequency scalings for all foreground components for each HEALPix pixel of \texttt{nside=1} which we then use to produce simulated maps for our experiments.
We call these simulations \texttt{\textit{customized d7s1 with spatial variability}} as they include both non parametric scaling and with spatial variability though limited again to the HEALPix \texttt{nside=1} pixels. 
While feasible, cases with more spatially varying scalings would require significant additional computational cost which we consider unwarranted at this stage.
For the \texttt{\textit{customized d7s1 with spatial variability simulations}}, we perform two runs, both for the satellite experiment and the input value of $r_{\rm{true}}=0$.
First, we run \texttt{MICMAC} adopting a single mixing matrix, $\mathbf{B}_{\mathbf{f}}$, for the entire observed sky. This gives a recovered value of $r$ of $r = (13.7 \pm 1.0) \times 10^{-3}$ ($95\%$ C.L.), thus showing a significant bias.
We then perform the same run allowing for a different mixing matrix for each \texttt{nside=1} HEALPix pixel. This gives $r < 0.4 \times 10^{-3}$ ($95\%$ C.L.) now compatible with $r_{\rm{true}}=0$.
In Fig.~\ref{fig:multipatch_non_parametric_r} we show the histograms of all the samples of $r$ drawn for each of the two runs, and in Fig.~\ref{fig:multipatch_non_parametric_Cls} the recovered power spectra for the residuals after component separation.

\section{Discussion and extensions}

We presented in this work a pixel domain version of the novel component separation approach first proposed in~\cite{leloup2023nonparametric}, and its implementation in a software package, \texttt{MICMAC}~\cite{MICMACgithub}. The former work was motivated by the need to develop a robust maximum likelihood component separation formalism handling Galactic foreground non-parametric spectral energy densities, but the proposed implementation could not handle spatial variability of the foreground SEDs. The pixel domain approach proposed here, can tackle foregrounds with spatial variability and non-parametric scaling laws. 

The ability to account for both such features is crucial for future CMB experiments targeting $B$ modes, as both are already known to be present in the actual foregrounds~\cite{krachmalnicoff2018s,akrami2020planck}. 
Importantly, the developed method is also naturally capable of dealing with the inhomogeneous noise properties, what is however out of scope of the present work.

This pixel domain variant of the method introduced here relies on a Gibbs sampling composed of four sampling steps. Two of the variables sampled are the CMB map and cosmological parameter, or alternatively CMB covariance, and rely on the established formalism for power spectrum estimation with Gibbs sampling. 
The two others are the free mixing matrix elements and latent parameter. 
In this work we have detailed algorithmic and numerical techniques needed to implement all these steps.

The approach described in this work can be compared to other methods such as the Internal Linear Combination methods~\cite{ILC}, as was done in~\cite{leloup2023nonparametric}. The ILC method relies on the data covariance to characterize frequency weightings, while the former requires a fixed expectation of the CMB covariance. Both methods do not assume foregrounds SED but can be extended to use an additional prior knowledge. The constrained ILC~\cite{cILC} method for instance uses the knowledge of a contaminant SED to properly remove it from the observed data. The \texttt{MICMAC} approach can similarly be extended to use additional information in the mixing matrix by adding components with a known SED contaminant. 
An important difference to highlight in both methods are the number of parameters to estimate, being the number of frequencies $n_f$ for the ILC method and $(n_f - n)n$ for \texttt{MICMAC} with $n$ the number of foreground components. 
The handling of the spatial variability of the foreground SEDs is as well very different for both approaches. \texttt{MICMAC} is based on a multi-patch approach, and requires the choice of a multi-patch distribution, while the ILC methods can be extended in needlets domain~\cite{2009A&A...493..835D, 2012MNRAS.419.1163B, 2013MNRAS.435...18B, Carones:2022xzs} and as such requires the choice of needlets. It should be noted that the needlet approach is harmonic-based, and not as flexible as the pixel-based multi-patch approach. We highlight as well that the ILC approaches are cheaper to compute.

We test and validate our implementation on two, prototypical cases of ground-based and satellite experiments, assuming the full and cut-sky cases with progressively more involved foregrounds.

In particular, we demonstrate the multi-patch functionality of the code in the satellite case and foreground models with spatial variability and parametric SEDs, the \texttt{\textit{customized d1s1}} model, and as well with non-parametric and spatially varying SEDs, the \texttt{\textit{customized d7s1 with spatial variability}} model. 
We also demonstrate the ability of the method to handle non-Gaussian foregrounds. 

The results show efficient foreground cleaning performance across all these different cases. Future work will evaluate the complete performances of the \texttt{MICMAC} method. This work will aim at assessing the limitations of the approach first against realistic foregrounds, but as well to test the flexibility of the method against for instance more realistic noise or an unanticipated additional foreground. 

The applicability of the approach in temperature component separation is as well left for a future work. 

The current formalism relies on a Gaussian description of the lensing, by neglecting the associated tiny non-Gaussianities. 
This approach can be tested further by possibly considering the non-Gaussianities as an addition to the Gaussian anisotropies, or by considering a more involved on-the-fly delensing approach to the Gibbs steps. 
An exploratory work is necessary to assess which approach is more efficient, and will be the subject of a future work. 

The current package is computationally heavy mostly due to time-consuming iterative linear systems solvers (PCG) and need for numerous spherical harmonic transforms (SHT). These steps can be readily improved on by using better, more specialized preconditioners for the PCG solvers ~\cite{Seljebotn_2019,MAPPRAISER_2022A&C....3900576E}, as well as better \texttt{JAX}-based spherical harmonics transform routines, \texttt{s2fft} package~\cite{price:s2fft}.
The work is on-going to improve the package with these features.

Finally, the formalism is well suited for a straightforward addition of instrumental effects, such as beams~\cite{Rizzieri_2024}, as well as for a possible extension to include more realistic noise covariances. Future work will address such effects. \\

\begin{acknowledgments}
The authors would like to thank Simon Biquard for his helpful comments leading to significant improvements in the implementation, Alexandre Boucaud for his valuable help in improving the quality of the package, Pierre Chanial for his help and contributions to the \texttt{JAX} tools used in \texttt{MICMAC}, notably with \href{https://github.com/CMBSciPol/jax-healpy}{Jax-Healpy}, and Gabriel Ducrocq for his useful comments in the early stages of the development of the method. We also thank the whole SCIPOL team for their continuous support and feedback throughout the development of the package, in particular Benjamin Beringue, Wassim Kabalan, Ema Tsang King Sang and Amalia Villarrubia-Aguilar. 
The authors acknowledge the use of the HEALPix~\cite{Healpix} and \texttt{healpy} packages \cite{Zonca2019,Healpy}, the \texttt{Numpyro} package \cite{phan2019composable,bingham2019pyro} and the \texttt{JAX} package~\cite{jax2018github, jax_paper}. 
JE acknowledges the SCIPOL project~\cite{Scipol} funded by the European Research Council (ERC) under the European Union’s Horizon 2020 research and innovation program (PI: Josquin Errard, Grant agreement No. 101044073). This work was granted access to the HPC resources of IDRIS under the allocation 2024-102865 made by GENCI. This work has also received funding by the European Union’s Horizon 2020 research and innovation program under grant agreement no. 101007633 CMB-Inflate. 
\end{acknowledgments}


\bibliographystyle{apsrev4-1}
\bibliography{bib_shortened}
\end{document}